%% file: 2008_bicep_cmb_2yr.tex
\documentclass{emulateapj}
\usepackage{apjfonts}
\usepackage{natbib}
\usepackage{graphics}
\usepackage{multirow}
\usepackage{amsbsy}
\usepackage{mathrsfs}
\input{defs}

\shorttitle{CMB POLARIZATION SPECTRA FROM BICEP TWO-YEAR DATA}
\shortauthors{CHIANG ET AL.}
\submitted{Astrophysical Journal 711 (2010) 1123--1140}

\begin{document}

\title{Measurement of Cosmic Microwave Background Polarization Power
  Spectra from Two Years of BICEP Data}

\author{H.~C.~Chiang\altaffilmark{1,2}}
\author{P.~A.~R.~Ade\altaffilmark{3}}
\author{D.~Barkats\altaffilmark{1,4}}
\author{J.~O.~Battle\altaffilmark{5}}
\author{E.~M.~Bierman\altaffilmark{6}}
\author{J.~J.~Bock\altaffilmark{1,5}}
\author{C.~D.~Dowell\altaffilmark{5}}
\author{L.~Duband\altaffilmark{7}}
\author{E.~F.~Hivon\altaffilmark{8}}
\author{W.~L.~Holzapfel\altaffilmark{9}}
\author{V.~V.~Hristov\altaffilmark{1}}
\author{W.~C.~Jones\altaffilmark{1,2}}
\author{B.~G.~Keating\altaffilmark{6}}
\author{J.~M.~Kovac\altaffilmark{1}}
\author{C.~L.~Kuo\altaffilmark{10,11}}
\author{A.~E.~Lange\altaffilmark{1,5,15}}
\author{E.~M.~Leitch\altaffilmark{12}}
\author{P.~V.~Mason\altaffilmark{1}}
\author{T.~Matsumura\altaffilmark{1}}
\author{H.~T.~Nguyen\altaffilmark{5}}
\author{N.~Ponthieu\altaffilmark{13}}
\author{C.~Pryke\altaffilmark{12}}
\author{S.~Richter\altaffilmark{1}}
\author{G.~Rocha\altaffilmark{1,5}}
\author{C.~Sheehy\altaffilmark{12}}
\author{Y.~D.~Takahashi\altaffilmark{9}}
\author{J.~E.~Tolan\altaffilmark{10,11}}
\author{K.~W.~Yoon\altaffilmark{14}}

\altaffiltext{1}{Department of Physics, California Institute of Technology, 
  Pasadena, CA 91125, USA}
\altaffiltext{2}{Department of Physics, Princeton University, Princeton, NJ
  08544, USA}
\altaffiltext{3}{Department of Physics and Astronomy, University of
Wales, Cardiff, CF24 3YB, Wales, UK}
\altaffiltext{4}{National Radio Astronomy Observatory, Santiago, Chile}
\altaffiltext{5}{Jet Propulsion Laboratory, Pasadena, CA 91109, USA}
\altaffiltext{6}{Department of Physics, University of California at
San Diego, La Jolla, CA 92093, USA}
\altaffiltext{7}{SBT, Commissariat \`a l'Energie Atomique, Grenoble, France}
\altaffiltext{8}{Institut d'Astrophysique de Paris, Paris, France}
\altaffiltext{9}{Department of Physics, University of California at
Berkeley, Berkeley, CA 94720, USA}
\altaffiltext{10}{Stanford University, Palo Alto, CA 94305, USA}
\altaffiltext{11}{Kavli Institute for Particle Astrophysics and Cosmology (KIPAC), Sand Hill Road 2575, Menlo Park, CA 94025, USA}
\altaffiltext{12}{University of Chicago, Chicago, IL 60637, USA}
\altaffiltext{13}{Institut d'Astrophysique Spatiale, Universit\'e
Paris-Sud, Orsay, France}
\altaffiltext{14}{National Institute of Standards and Technology,
Boulder, CO 80305, USA}
\altaffiltext{15}{Sadly, Andrew Lange passed away shortly before the
  publication of this article.}

\begin{abstract}

Background Imaging of Cosmic Extragalactic Polarization (\bicep) is a 
bolometric polarimeter designed to measure the
inflationary \bmode\ polarization of the cosmic microwave background
(CMB) at degree angular scales.  During three seasons of observing at
the South Pole (2006 through 2008), \bicep\ mapped $\sim2\%$ of the
sky chosen to be uniquely clean of polarized foreground emission.
Here, we present initial results derived from a subset of the data
acquired during the first two years.  We present maps of temperature,
Stokes $Q$ and $U$, $E$ and $B$ modes, and associated angular power
spectra.  We demonstrate that the polarization data are
self-consistent by performing a series of jackknife tests.  We study
potential systematic errors in detail and show that they are
sub-dominant to the statistical errors.  We measure the \emode\ angular
power spectrum with high precision at $21 \leq \ell \leq 335$,
detecting for the first time the peak expected at $\ell \sim 140$.
The measured \emode\ spectrum is consistent with expectations from a
\lcdm\ model, and the \bmode\ spectrum is consistent with zero.  The
tensor-to-scalar ratio derived from the \bmode\ spectrum is $r =
0.02^{+0.31}_{-0.26}$, or $r < 0.72$ at 95\% confidence, the first
meaningful constraint on the inflationary gravitational wave
background to come directly from CMB \bmode\ polarization.

\end{abstract}


\section{Introduction}

One of the cornerstones in our current understanding of cosmology is
the theory of inflation.  Inflation addresses several major
shortcomings of the standard big bang model, resolving the flatness and
horizon problems and explaining the origin of structure; 
however, the theory has yet to be unambiguously confirmed by 
observational evidence.

Numerous experiments have demonstrated that the cosmic microwave
background (CMB) is an extremely effective tool for studying the early
universe.  Precision measurements of the temperature anisotropies now
span a wide range of angular
scales~\citep{jones06,reichardt09,nolta09,friedman09,sievers09,brown09} and
have yielded tight constraints on a model of the
universe in which the energy content is dominated by a cosmological
constant and cold dark matter (\lcdm).  

The polarization anisotropies
of the CMB provide even more insight into the history of the universe,
potentially encoding information from long before the moment of
matter--radiation decoupling.  The primary source of CMB polarization
is Thomson scattering of the local quadrupole of the photon--baryon
fluid sourced by density fluctuations.  The resulting partial
polarization has no handedness and is called the gradient or
``\emode '' by analogy to curl-free electric fields.  The \lcdm\
parameters that predict the temperature spectrum also predict the
shape of the \emode\ spectrum with almost no additional information
(reionization enhances power at the largest angular scales).  \emode\
polarization was first detected by \dasi~\citep{kovac02} and has since
been measured by many other
experiments~\citep{leitch05,montroy06,sievers07,wu07,bischoff08}.  The
acoustic peaks in the polarization spectra have been 
measured to high precision 
in both the $TE$ spectrum~\citep{piacentini06,nolta09} and, more
recently, directly in the $EE$ spectrum~\citep{pryke09,brown09}, providing
further support for our basic understanding of CMB physics.

Inflation predicts the existence of a stochastic gravitational wave
background, created during the initial accelerated expansion of the
universe, which imparts a unique imprint on the CMB at the surface of
last scattering~\citep{polnarev85}.  In addition to producing \emode\ polarization,
gravitational waves also induce a curl or ``\bmode'' in the
polarization anisotropies~\citep{seljak97,kamionkowski97}.  The inflationary 
\bmode\ signal is expected to peak at degree angular scales
(multipole moment $\ell\sim100$) with an amplitude determined by the energy scale of
inflation.  Because density fluctuations at the surface of last
scattering create only \emode\ polarization, a detection of the
\bmode\ signal would be strong evidence that inflation occurred 
\citep[see, e.g.,][]{dodelson09}.

The inflationary \bmode\ amplitude is parameterized by the tensor-to-scalar ratio
$r$, and the 
most restrictive published 
upper limit, $r<0.22$ (95\% confidence), comes
from measurements of large-scale temperature anisotropies in
combination with baryon acoustic oscillation and Type~Ia supernova
data~\citep{komatsu09}.  However, the constraints from temperature
anisotropies are ultimately limited by cosmic variance, and lowering
the $r$ limit further requires direct polarization measurements.
Currently, limits from polarization are still far worse than those
from temperature---for example, assuming \lcdm\ parameters that are
fixed at \wmap\ best-fit values, the \wmap\ \bmode\ spectrum
constrains $r<6$ at 95\% confidence.  The results reported in 
this paper provide upper limits on the \bmode\ signal that are an order of
magnitude more stringent than those set by \wmap.

Background Imaging of Cosmic Extragalactic Polarization (\bicep) is a
microwave polarimeter that has been designed specifically to probe the
\bmode\ of CMB polarization at degree angular scales.  The instrument
observed from the South Pole between 2006 January and 2008 December.
A detailed description of the \bicep\ instrument characterization
procedures is given in a separate paper, \citet{takahashi09}, that
accompanies this text.  In this paper, we report initial CMB
polarization results from the 2006 and 2007 observing seasons.

\section{The BICEP Instrument}

A complete description of the \bicep\ instrument is available
in~\citet{yoon06}, and only a brief summary is given here.  The
\bicep\ receiver consists of a two-lens refracting telescope coupled
to a focal plane of 49 orthogonal pairs of polarization-sensitive
bolometers \citep[PSBs;][]{2003SPIE.4855..227J}.  The PSB pairs are
divided between 25 that observe at 100~GHz and 24 at 150~GHz (two of
the 150 GHz PSB pairs were reconfigured for 220~GHz operation in late
2006 and were subsequently not used for CMB analysis).  The angular
resolution at 100 and 150~GHz is 0.93$\deg$ and 0.60$\deg$,
respectively, and the instantaneous field of view is 18$\deg$.  The
entire focal plane and optics assembly is housed in an upward-looking
cryostat with toroidal liquid nitrogen and liquid helium tanks.  The
clean optical path and azimuthal symmetry minimize instrumental
polarization systematics.

The receiver is supported in an azimuth--elevation mount with a third
degree of rotational freedom about the boresight.  The mount is
located on the top floor of the Dark Sector Laboratory (89.99$\deg$~S,
44.65$\deg$~W) at the Amundsen--Scott South Pole station, a site with
excellent atmospheric transparency and stability at millimeter
wavelengths as well as outstanding infrastructure.  The telescope
penetrates through the roof and is sealed to the building with a
flexible environmental enclosure, leaving most of the instrument
accessible in a warm lab setting.

\begin{figure}[t]
\resizebox{\columnwidth}{!}{\includegraphics{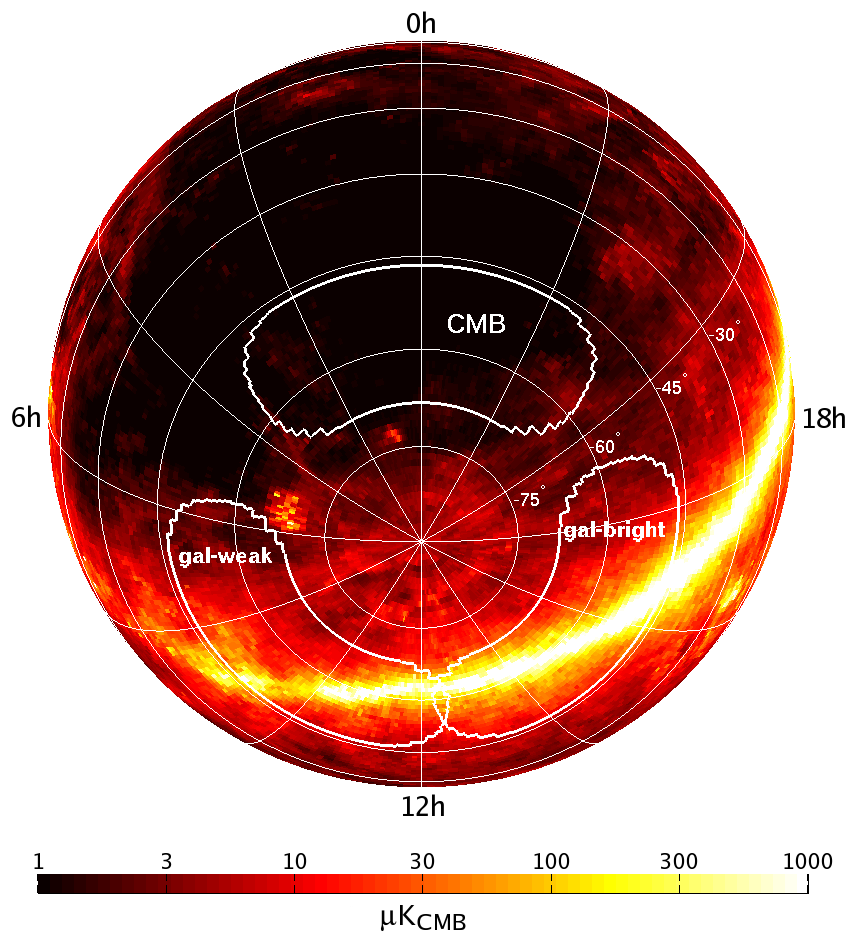}}
\caption{\bicep's CMB and Galactic fields are outlined on the 150-GHz
FDS Model~8 prediction of dust emission~\citep{finkbeiner99}, plotted 
here in equatorial coordinates. \\}
\label{fig:coverage_fds}
\end{figure}

The 24-hr visibility of the target field from the South Pole enables
uninterrupted observation and deep integration.  \bicep's primary CMB
field lies within the ``Southern Hole,'' a region of low dust emission
outlined in Figure~\ref{fig:coverage_fds}, in a right ascension and
declination range of approximately $|\alpha| < 60\deg$ and $-70\deg
< \delta < -45\deg$.  The telescope observation cycle is 48 sidereal
hours in length and is divided into four 9-hr CMB observations,
6~hr of Galactic observations, and 6~hr of cryogen
servicing.  The CMB field is covered twice over the same azimuth range
during each 48-hr cycle, but the elevation halves are mapped in
opposite order between the two observations.  The boresight angle in
each cycle is held fixed at one of four angles $\{-45\deg, 0\deg, 180\deg,
135\deg\}$ that provide good thermal-microphonic stability
and redundant polarization angle coverage.

\bicep\ maps the sky with azimuth--elevation raster scans.  During
each complete CMB observation (18 hr), the telescope boresight
steps in elevation between 55$\deg$ and 59.75$\deg$ in 0.25$\deg$
increments.  At each step in elevation, the telescope performs a set
of 50 back-and-forth azimuth scans over a total period of $\sim$50
minutes.  The azimuth scan width is 64.4$\deg$, and the speed is held
constant at 2.8$\deg$/s over $\sim$70\% of the scan duration, which
modulates the sky signal and places it in a frequency band of
approximately 0.1--1~Hz.  The scans have a fixed azimuth center that
is updated at each elevation step to approximately track the field center.  This
scan strategy was chosen instead of continuous tracking in order to
allow removal of any azimuth-fixed or scan-synchronous contamination.

Relative detector gains are measured regularly during observing cycles
with ``elevation nods'' performed at the beginning and end of each
fixed-elevation scan set.  During an elevation nod, the mount 
performs a rounded triangle wave motion in elevation with a 1.2$\deg$
peak-to-peak amplitude, and the detector voltages vary in response to
the changing line-of-sight air mass.  The nod is performed over a 45-s
period to reduce thermal disturbances on the focal plane, and thermal
drifts are further suppressed by using mirror-image elevation nods at
the beginning and end of each scan set (up-down-up and down-up-down).

\section{Instrument Characterization}\label{s:characterization}

The timestream $d(t)$ of a perfect linearly polarized detector is
related to the signal on the sky through the expression $d(t) =
T({\bf r}_b) + Q({\bf r}_b)\cos2\psi + U({\bf r}_b)\sin2\psi$, where the sky
signal is described by the Stokes parameters $T, Q, U$.  This
expression depends only on the detector's direction of observation ${\bf r}_b$ and
 polarization orientation angle $\psi$.  (The
time-dependence of ${\bf r}_b$ and $\psi$ that arises from the scan
strategy is suppressed for clarity.)
Modifying this simple expression, the timestream response of a \bicep\
PSB is described by
\begin{eqnarray}
\nonumber d(t)&=&K(t)\ast{\Big\{}n(t)+g(t)\int d\nu A_e F(\nu) 
\int d{\bf r} \: {\cal P}({\bf r}-{\bf r}_b,\nu) \\
&& [ T({\bf r},\nu) + 
{{1-\epsilon} \over {1+\epsilon}} (Q({\bf r},\nu)\cos 2\psi + 
U({\bf r},\nu)\sin 2\psi) ] {\Big\}},
\label{eq:full_bolo}
\end{eqnarray}
where the extra terms are calibration quantities that describe the
properties of the instrument.  The cross-polar leakage, which is a
PSB's level of response to orthogonally polarized light, is
parameterized by $\epsilon$.  The detector signal is convolved with
the co-polar beam ${\cal P}({\bf r},\nu$), which has a spatial extent that depends
on the coordinate ${\bf r}$.  The beam-convolved detector response is
integrated over the pass band $F(\nu)$, which is multiplied by the
effective antenna area $A_e$.  The gain factor $g(t)$ converts voltage
to temperature units, and $n(t)$ is an additive noise term.  Finally,
the entire expression is convolved with the detector transfer function
$K(t)$.

In order to faithfully reconstruct the temperature and polarization signal on the sky,
it is necessary to measure all the terms in
Equation~\ref{eq:full_bolo} that relate $T, Q, U$ to the detector
voltage.  A complete description of these measurement procedures and
results is given in the accompanying instrument characterization
paper~\citep{takahashi09}.  Here, we give only a brief summary of the
quantities used directly in data analysis.  The list includes detector
transfer functions, absolute and relative gains, main beam shapes,
cross-polar leakages, detector orientation angles, and pointing.  The
noise is discussed separately in \S\ref{s:nsim}.

\subsection{Transfer Functions}

Deconvolving detector transfer functions is the first step in
producing clean timestreams that are suitable for analysis.  Relative
gains are measured with elevation nods at 0.02~Hz, so the transfer
functions must be characterized over a frequency range that spans at
least 0.01--1~Hz in order to link the relative gains to the entire
science band.  The transfer functions were measured with a microwave
source (Gunn oscillator or broadband noise source) that was placed
near the telescope window and square-wave modulated at 0.01~Hz.
The time-domain responses to the transitions were Fourier transformed,
divided by the transform of the modulation waveform, and averaged for
each detector in order to obtain the deconvolution kernel.
The transfer functions have a measurement precision of $0.5\%$ rms
across the signal band and have sufficiently high signal-to-noise ratio (S/N) to
be deconvolved directly from the bolometer timestreams.  (Although the
transfer functions can be described by a simple model, we do not rely
on those fits.)  The relative gain
uncertainty that results from transfer function deconvolution is
$<0.3\%$ over the range 0.01--1~Hz.

\subsection{Relative and Absolute Gains}\label{s:gains}

PSB relative gains are measured with elevation nods at the beginning
and end of every set of constant-elevation azimuth scans.  The nods inject an atmospheric
signal into the bolometer timestreams, and the gains are
obtained by fitting each timestream against the cosecant of
the detector elevation.  The resulting volts-per-air mass responsivity
factors are normalized to the average of good detectors for each
frequency band during the scan set.  The common-mode rejection of CMB
temperature anisotropies in gain-adjusted PSB pairs is measured to be
better than $98.9\%$ at degree angular scales.

We cross-correlate the CMB temperature fluctuations measured by \wmap\
and \bicep\ to obtain absolute gains, which relate CMB temperature to 
detector units.  The \wmap\ maps are smoothed to \bicep's resolution by
applying the ratio of the beam window functions, $B^{\rm
BICEP}_\ell / B^{\rm WMAP}_\ell$.  (We do not apply a correction for the pixelization 
of the \wmap\ maps since the effects are negligible.)  The smoothed maps are then
converted to simulated detector timestreams using the boresight pointing data.
The timestreams are filtered and converted back into maps, thus
creating a ``\bicep-observed'' version of the \wmap\ data.  The
\bicep\ map and processed \wmap\ maps, which have compatible beam and
filter functions, are cross-correlated in multipole space to obtain
the absolute gain
\begin{equation}
g_b = { \sum_{\ell} P_\ell^b {\langle a^{\rm WMAP-1}_{\ell m} {a^{*\: {\rm BICEP}}_{\ell
m}} \rangle} \over {\sum_{\ell} P_\ell^b \langle a^{\rm WMAP-1}_{\ell m} a^{*\: {\rm
WMAP-2}}_{\ell m} \rangle} },
\label{eq:gain_cal}\end{equation}
where $P_\ell^b$ is a top hat binning operator, and $a_{\ell m}$ are 
the spherical harmonic expansion coefficients.
To avoid noise bias, the $a_{\ell m}$ coefficients in the denominator
are taken from two different \wmap\ maps; for this analysis, we have
used the Q and V band maps from the five-year data release~\citep{hinshaw09}.  The
resulting gain calibration, $g_b$, is approximately flat over
\bicep's $\ell$~range of 21--335, where the lower bound is set by the
timestream filtering, and the upper bound is set by beam uncertainty.
The absolute gain used for each of the \bicep\ frequency bands is a
single number taken from the average of $g_b$ over six uniform bins spanning 
a multipole window
of $56 \le \ell \le 265$, and the absolute gain uncertainty is derived
from the standard deviation of $g_b$.  To assess the impact of
errors in the beam window functions, we have calculated $g_b$ using
both the Q and V band \wmap\ data in the numerator of
Equation~\ref{eq:gain_cal}.  The average $g_b$ values are
consistent within errors, and we take the larger of the two 
standard deviation values, 2\%, as a conservative estimate of the
absolute gain uncertainty in temperature units.

\subsection{Boresight and Detector Pointing}

The two components of detector pointing reconstruction are telescope
boresight pointing and detector offsets relative to the boresight
(focal plane coordinates).  Raw boresight pointing timestreams are
obtained from encoders located on the three mount axes, and
corrections to the raw data are applied from a model describing axis
tilts and encoder offsets.  The pointing model is established with
star observations from an optical camera located on the upper surface
of the cryostat.

The focal plane coordinates of each detector are reconstructed from
measurements of CMB temperature fluctuations.  Temperature maps from
each detector at each of the four boresight angles 
are cross-correlated with the full-season map, and the
process is repeated with varying centroid adjustments until the maps
are self-consistent.  This method results in $0.03\deg$ rms centroid
uncertainty.

\subsection{Polarization Orientation and Efficiency}

To recover polarization information from detector timestreams, it is
necessary to know the polarization orientation angle $\psi$ and
cross-polar leakage $\epsilon$ of each PSB.  Note that $\epsilon$ is a
property of the detector itself and is independent of the cross-polar
beam, which is a property of the optical chain upstream of the PSB.
For \bicep, $\epsilon$ dominates over the cross-polar beam.  Both
$\psi$ and $\epsilon$ are measured with several different devices
(e.g.,\ rotating wire grid, dielectric sheet calibrator) that send
polarized light into \bicep\ at many different angles with respect to
the detectors.  The phase of each PSB's sinusoidal response and the
ratio of the minimum to maximum determine $\psi$ and $\epsilon$,
respectively.  The uncertainty in the measured orientation angles is
$\pm0.7\deg$.  The median $\epsilon$ for \bicep\ PSBs is about $0.04$,
with a measurement uncertainty of $\pm0.01$.

\subsection{Main Beam Shapes}

The \bicep\ beams were mapped by raster scanning the telescope over a
bright source at various fixed boresight angles.  The beams are well
described by a Gaussian model, with fit residuals typically about
1\% with respect to the beam amplitude.
The beams at both frequencies are nearly circular, with ellipticities
under 1.5\%.  We
therefore approximate the beams as symmetric Gaussians with average
full widths of 0.93$\deg$ and 0.60$\deg$ at 100 and 150~GHz,
respectively.  The distribution of beam widths varies by $\pm3\%$
across the focal plane, and each width is measured to a precision of
$\pm0.5\%$.

\section{Low-level Analysis and Mapmaking}

The analysis presented here includes \bicep\ data from the 2006 and
2007 observing seasons.  For these initial results, we restrict the
data set to uninterrupted 9-hr CMB observations taken during
February--November.  Although there is no evidence for Sun
contamination, we exclude data acquired during the Austral summer
because of mediocre weather conditions and increased station
activities.  A significant fraction of 2006 was devoted to calibration
measurements and investigation of a different scan strategy, and the
nominal data set in this season consists of 124 days.  The amount of
data increased to 245 days in the 2007 season.

\subsection{Data Cuts}

The data set is further reduced by omitting 9-hr observation
phases with extremely poor weather quality.  For each phase, the
standard deviation of relative gains from elevation nods is calculated
for each channel, and the median over the 100 and 150-GHz channels of
those standard deviations yields two numbers per phase.  An
observation phase is cut if either of the median standard deviations
is greater than 20\% of the average relative gain.  After this weather
cut, the data set consists of 117 and 226 days for the 2006
and 2007 seasons.

Of the 49 optically active PSB pairs, several are excluded from each
season due to anomalous behavior such as no response, excess noise,
poorly behaved or time-dependent transfer functions, and exceptionally
poor polarization efficiency.  Six experimental pixels containing
Faraday rotator modules (2006 season only) and the two 220-GHz pixels
(2007 and 2008 seasons) are also excluded from CMB analysis.  A
total of 19 100-GHz and 14 150-GHz PSB pairs are used for 2006
analysis, and 22 100-GHz and 15 150-GHz PSB pairs are used for 2007.

\subsection{Low-level Timestream Processing}

The raw output of \bicep\ consists of voltage timestreams sampled at
50 Hz for 144 channels comprising 98 light bolometers, 12 dark
bolometers, 16 thermistors, 10 resistors, and 8 intentionally open
channels.  The low-level timestream cleaning begins with concatenating
and trimming the raw data files for each 9-hr observation phase.
Complete transfer functions are deconvolved for the 98 light
bolometers, and all timestreams are low pass filtered at 5~Hz and
downsampled to 10~Hz.  Relative detector gains are derived from
elevation nods, and horizon and celestial boresight coordinates are
calculated using the pointing model.  ``Half-scans,'' or single sweeps
in azimuth, are identified, and the turnarounds are excluded from
further analysis.  The remaining central portion of the half-scan,
which has nearly constant velocity, makes up $\sim75\%$ of the total
scan duration.

Each detector half-scan is subjected to three data quality checks.
First, relative gains derived from elevation nods at the beginning and
end of a set of scans are compared, and all of the half-scans in the
scan set are excluded if the gains differ by more than 3\%.
Second, gain-adjusted PSB pairs are differenced over each half-scan,
and the half-scan is rejected if the magnitude of the skew or kurtosis
is abnormally high (PSB pair-difference timestreams are expected to
be Gaussian white noise).  Finally, half-scans that contain
cosmic rays and other signal spikes are identified by 
points that are greater than 7 times the standard deviation of the
smoothed timestream.  On average, the combined half-scan flagging
criteria exclude about 3\% of all half-scans over all light detectors.
Because the flagged percentage is small, the problematic half-scans
are not gap-filled and are simply omitted from analysis.

\subsection{Mapmaking}

\begin{figure*}[h]
\resizebox{\textwidth}{!}{\includegraphics{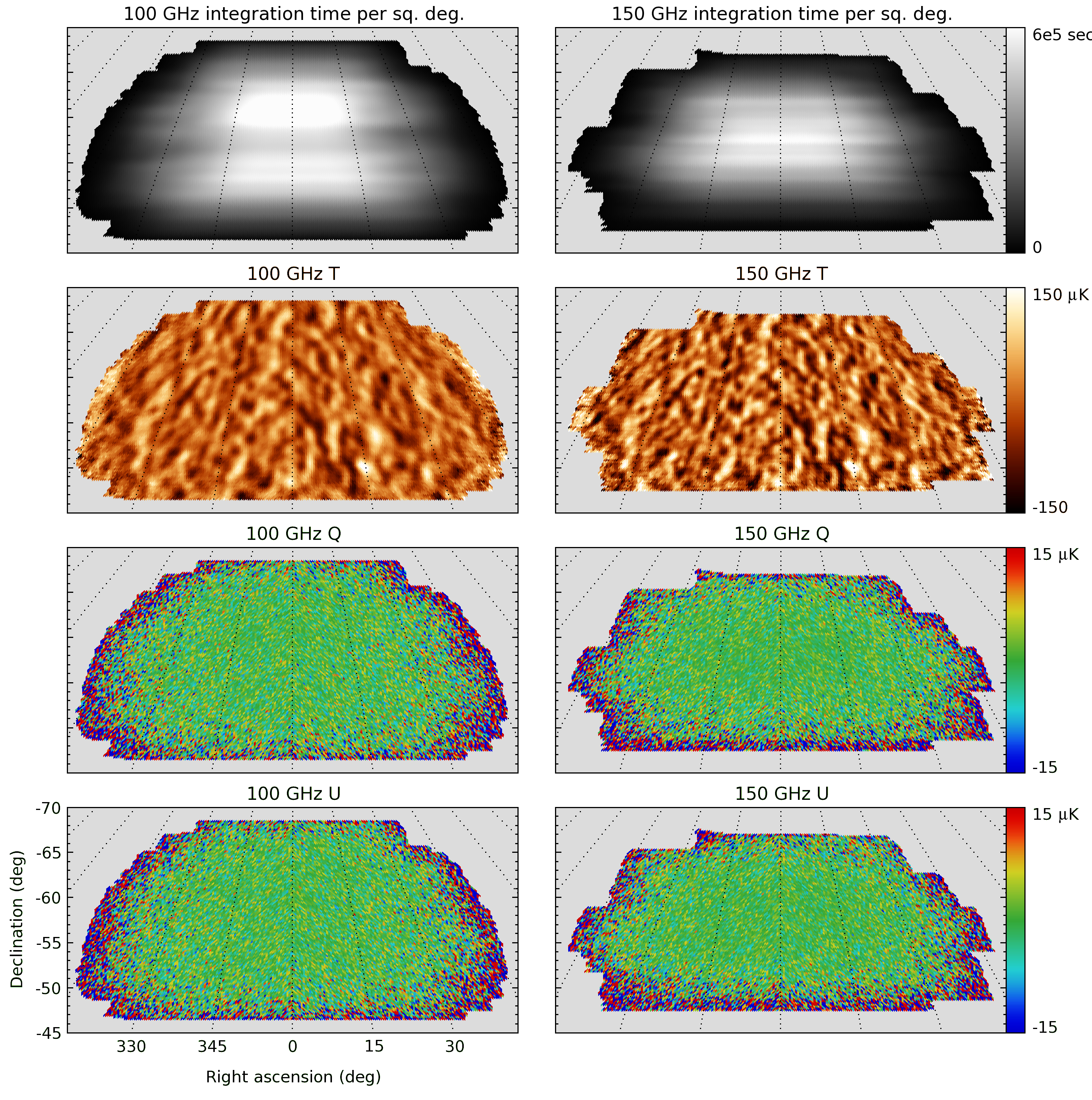}}
\caption{\bicep\ $T$, $Q$, $U$, and coverage maps.  The resolution is
about 0.9$\deg$ and 0.6$\deg$ at 100 and 150~GHz, respectively, and no
smoothing or apodizing has been applied to the maps.  The noise per
square degree in the central region of the $Q$ and $U$ maps is
0.81~$\mu$K at 100~GHz and 0.64~$\mu$K at 150~GHz.  Note that the
color scales of the temperature and polarization maps differ by a
factor of 10.}
\label{fig:tqu_maps}
\end{figure*}

After low-level cleaning, the bolometer timestreams are binned into
temperature and polarization maps.  We have developed two data
analysis pipelines for \bicep\ that differ starting from the
mapmaking stage.  The code in each pipeline is completely independent
of the other, and the only shared data products are the initial set of
downsampled, cleaned detector timestreams, boresight pointing, and
calibration data.
Both pipelines have reproduced all the results reported in this paper,
achieving similar statistical power and excellent agreement.
The mapmaking algorithms used by the two pipelines are similar,
although one (designated for this paper the ``primary pipeline'') 
bins in the \healpix~\citep{gorski05} pixelization scheme,
while the other (designated for this paper the ``alternate pipeline'')
produces maps in rectangular coordinates.  In this section, we describe
the primary pipeline's mapmaking procedure in detail.

Following ~\citet{jones07},
the simplified timestream output $d_{ij}$ of a single PSB can be
expressed as
\begin{equation}
d_{ij} = g_{ij} \left[T({\bf p}_j) + \gamma_{i} (Q({\bf p}_j)\cos2\psi_{ij} + 
U({\bf p}_j)\sin2\psi_{ij})\right],
\label{eq:psb_timestream}
\end{equation}
where $g$ is the gain, $T, Q, U$ are the beam-integrated Stokes
parameters of the sky signal, $\gamma \equiv (1 - \epsilon) / (1 +
\epsilon)$ is the polarization efficiency factor, and $\psi$ is the
PSB polarization orientation projected on the sky.  The index $i$
denotes the PSB channel number, $j$ is the timestream sample number,
and ${\bf p}_j$ is the map pixel observed at time $j$.  The goal of
mapmaking is to recover $T, Q, U$ from the bolometer timestreams.

The mapmaking procedure for \bicep\ 
begins with the formation of gain-adjusted sum and difference 
timestreams for each PSB pair:
\begin{equation}
d^\pm_{ij} = {1\over2} (d_{2i,j} / g_{2i,j} \pm d_{2i+1,j} / g_{2i+1,j}).
\label{eq:psb_sumdif}
\end{equation}
To reduce atmospheric $1/f$ noise, a third order polynomial is
subtracted from the sum and difference timestreams for each half-scan
in azimuth.  Azimuth-fixed and scan-synchronous contamination are
removed by subtracting a template signal, which is formed by binning
the polynomial-filtered detector timestreams in azimuth over each set of fixed-elevation
scans.  There are slight differences in the scan-synchronous signal between
left- and right-going half-scans, so separate templates are
calculated for each case.
The scan-synchronous contamination removed in this step is very
small; $Q$ and $U$ maps typically change by 100--400~nK at the
largest scales after its removal.

The temperature
$T$ at each map pixel ${\bf p}$ is recovered from the filtered sum
timestreams with
\begin{equation}
\left( \sum^{\rm No. of pairs}_i \sum_{j \in {\bf p}} w^+_{ij} d^+_{ij} \right) {\Bigg /}
\left( \sum^{\rm No. of pairs}_i \sum_{j \in {\bf p}} w^+_{ij} \right)
\simeq T({\bf p}),
\end{equation}
where $w^+$ is the weight assigned to each pair sum, and we have
assumed that the effects of polarization leakage are negligible.  In
other words, the temperature is obtained simply by binning the
filtered detector timestreams into map pixels.  Stokes $Q$ and $U$ are
calculated from linear combinations of the difference timestreams
using the matrix equation
\begin{eqnarray}
\nonumber && \sum^{\rm No. of pairs}_i \sum_{j \in {\bf p}} w^-_{ij}
\left(\begin{array}{c}
d^-_{ij} \alpha_{ij} \\
d^-_{ij} \beta_{ij} \\
\end{array}\right) = \\
&& {1\over2} \sum^{\rm No. of pairs}_i \sum_{j \in {\bf p}} w^-_{ij}
\left(\begin{array}{cc}
\alpha^2_{ij} & \alpha_{ij}\beta_{ij} \\
\alpha_{ij}\beta_{ij} & \beta^2_{ij}\\
\end{array}\right)
\left(\begin{array}{c}
Q({\bf p}) \\
U({\bf p}) \\
\end{array}\right).
\label{eq:qu_diff}\end{eqnarray}
Here, $w^-$ is the weight assigned to each pair difference, and
$\alpha, \beta$ are PSB pair orientation angle factors defined as
\begin{eqnarray}
\alpha_{ij} = \gamma_{2i}\cos2\psi_{2i,j} - \gamma_{2i+1}\cos2\psi_{2i+1,j} \\
\beta_{ij} = \gamma_{2i}\sin2\psi_{2i,j} - \gamma_{2i+1}\sin2\psi_{2i+1,j}.
\end{eqnarray}
The $2\times2$ matrix in Equation~\ref{eq:qu_diff} is singular for a
single pair of $\psi_{2i,j}$ and $\psi_{2i+1,j}$, and the equation can
be solved only by accumulating more than one timestream sample in a
given map pixel ${\bf p}$.  As ${\bf p}$ is observed with many
detector polarization angles $\psi$, the off-diagonal
$\alpha_{ij}\beta_{ij}$ terms average to zero, and the matrix becomes
invertible.  Although only two different polarization angles are
required to invert the matrix, some instrumental systematics average
down as the number of observation angles increases.  By examining the
determinant of the matrix, pixels (typically at the edge of the map)
with insufficient polarization angle coverage or low integration time
are identified and masked from analysis.

We choose the pair sum and difference weights $w^\pm$ to be
proportional to the inverse variance of the filtered timestreams.
The weights are evaluated from power spectral densities averaged over
each set of azimuth scans (every 50 minutes), a period during which
the noise properties are approximately stationary.  For each channel
pair, the sum/difference weight for a scan set is calculated from the
inverse of the average value of the autocorrelation between 0.5 and 1~Hz.

\section{Map Results}

\begin{figure*}
\resizebox{\textwidth}{!}{\includegraphics{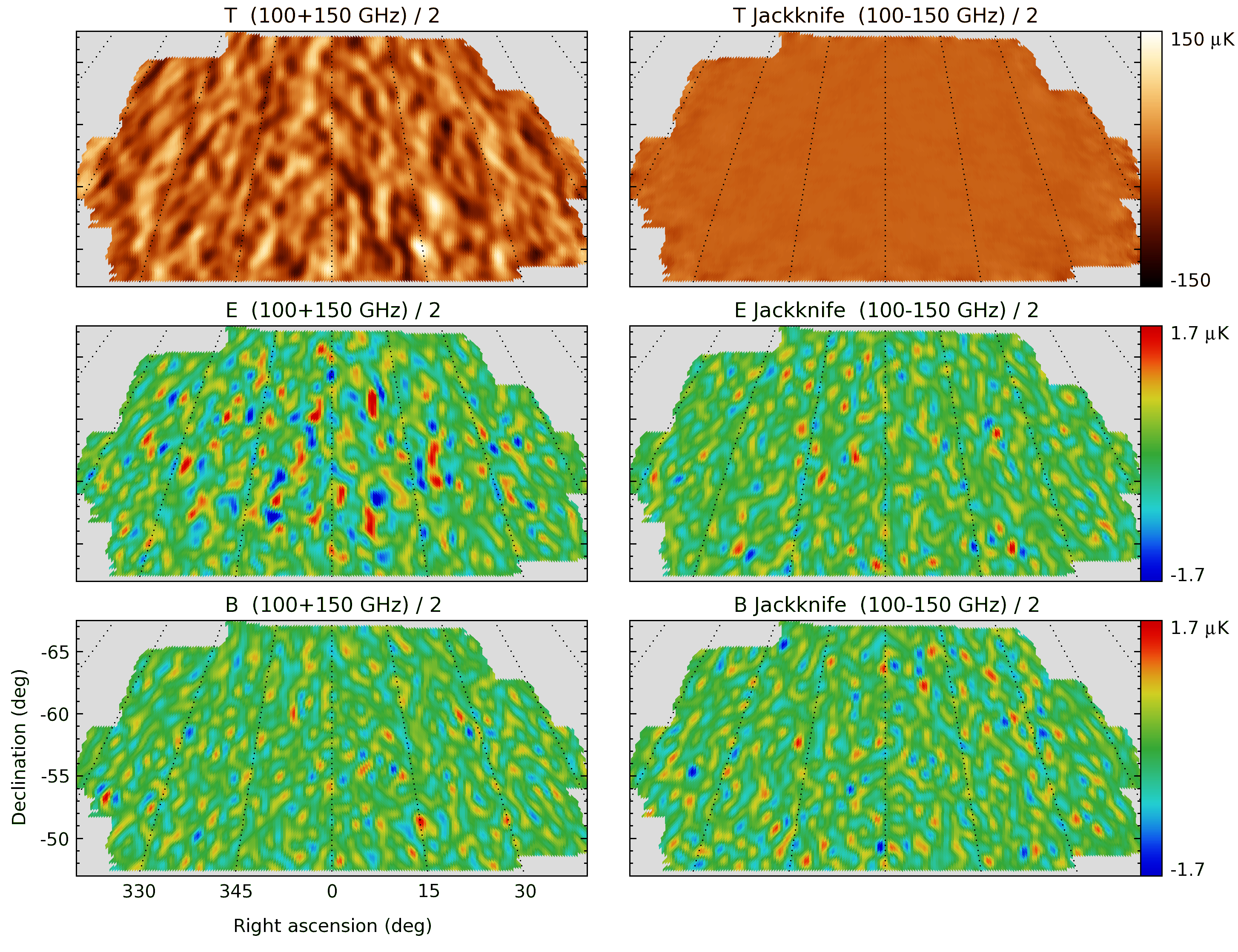}}
\caption{Data from \bicep's 100-GHz and 150-GHz channels are combined
to form temperature, $E$, and $B$ signal and jackknife maps.  The $E$
and $B$ maps are apodized to downweight noise-dominated edge pixels.  The
temperature anisotropies are measured with high S/N, and
the $E$ signal map shows resolved degree-scale structure.  The $B$
signal map and the $E$ and $B$ jackknife maps are consistent with
noise. \\}
\label{fig:teb_maps}
\end{figure*}

Figure~\ref{fig:tqu_maps} shows the \bicep\ maps of CMB temperature
and Stokes $Q$ and $U$ parameters.  The pixelization is 0.23$\deg$,
corresponding to \healpix\ nside=256, and the total observation area
covers $\sim2\%$ of the sky.  In the central part of the maps, the
integration time exceeds $6\times10^5$ detector--seconds per square
degree, and the scan strategy smoothly apodizes the outer edges of the
observed region.  The temperature anisotropies are measured with high
S/N and agree well between the two frequencies.  The rms
noise per square degree, obtained from simulations, is 0.81~$\mu$K and
0.64~$\mu$K for 100~GHz and 150~GHz, respectively, in the central
region of the $Q$ and $U$ maps.

Data from both frequency bands are combined to form the temperature, $E$, and
$B$ maps shown in Figure~\ref{fig:teb_maps}.  The maps from each frequency
band are smoothed to 1$\deg$ resolution, and the left column shows the sum.
Frequency jackknife maps that are formed by differencing the 100 and 150~GHz
data are shown in the right column.  The high S/N of the
temperature measurement is demonstrated by the lack of structure in the
frequency jackknife map.  The faint striping, which is caused by residual
unpolarized atmospheric noise, is successfully removed by PSB differencing.  To form the
$E$ and $B$ maps, the $a^E_{\ell m}$ and $a^B_{\ell m}$ coefficients are
computed from apodized $Q$ and $U$ maps using the \anafast\ utility in the
\healpix\ code package.  The coefficients are then boxcar
filtered in a multipole range of $70<\ell<280$.  Using the \synfast\ utility,
$a^{E,B}_{\ell m}$ are interpreted as $Y_{\ell m}$ coefficients from which the
$E$ and $B$ maps are generated.  The $E$ frequency-sum map shows resolved
degree-scale structure of the expected amplitude, while the $E$ and $B$
jackknife maps, as well as the $B$ signal map, are consistent with noise.
These $E$ and $B$ maps are used only as a qualitative illustration of
the S/N ratio of the data.  Since the method of creating
the maps has several shortcomings, e.g.\ there is some $E$/$B$ mixing
due to the finite survey area, we do not use the $E$ and $B$ maps in
any quantitative analysis of the power spectra.

\section{Power Spectrum Estimation}\label{s:ps_estimation}

\bicep\ employs two different analysis pipelines for power spectrum
estimation: the first performs a curved-sky analysis using the
publicly available \spice~\citep{chon04} package to estimate the power
spectra, and the second is a flat-sky pseudo-$C_\ell$ estimator
adapted from the \QUAD\ pipeline that was used in the \citet{pryke09} main analysis.  
In the following,
we describe in detail the \spice\ pipeline.

The power spectra are derived from inverse variance weighted 
$T$, $Q$, and $U$ maps.  Temperature and polarization
correlation functions are computed from fast spherical harmonic transforms 
of the heuristically weighted maps.
The polarization correlation functions are then decoupled so that
$E$/$B$ mixing that is caused by the finite survey area is removed in the
mean.  The correlation functions are apodized 
in order to reduce correlations in multipole space that result from 
incomplete sky coverage.  Finally, estimates of the full sky power spectra are
computed from integral transforms of the apodized correlation functions.  
The observed power spectra, as computed by \spice, are approximated as
\begin{equation}
\hat{C}_\ell^{X} = \sum_{\ell'} \kappa^X_{\ell\ell'} F^X_{\ell'} {\cal
 B}_{\ell'}^2 C_{\ell'}^{X} +\hat{N}_\ell^X,
\label{eq:psmodel}
\end{equation}
where the superscripts, $X$, correspond to the six
temperature--polarization combinations $\{TT$, $EE$, $BB$, $TE$, $TB$,
$EB\}$.  The convolution kernel, $\kappa^X_{\ell\ell'}$, is the normalized
\spice\ window function and depends on the apodization function
applied to the correlation functions.  The effect of timestream
filtering is described by $F^X_\ell$, the $\ell$-space transfer
function; ${\cal B}_\ell^2 = B_\ell^2 H_\ell^2$ is the product of the
beam and pixel window functions, and $\hat{N}^X_\ell$ is the power
spectrum of the noise convolved with the \spice\ kernel.  In the
following subsections, we describe the steps in solving
Equation~\ref{eq:psmodel} and recovering an estimate of the
underlying power spectrum, $C^X_\ell$.

\subsection{Noise Subtraction}\label{s:nsim}

The first step in power spectrum estimation is calculating and
subtracting the noise bias ${\hat N}_\ell$.  We estimate ${\hat
N}_\ell$ with Monte Carlo simulations of instrument noise: starting
from a noise model, we generate simulated timestreams that are
filtered, co-added into maps, and processed with \spice.  The resulting
simulated noise spectra are averaged over many realizations to form
${\hat N}_\ell$.

The \bicep\ noise model is derived from gain-adjusted PSB pair sum and
difference timestreams (Equation~\ref{eq:psb_sumdif}) under the
assumption that the timestream S/N is negligible,
i.e., the signal is the noise.  The S/N is $\leq10\%$ and $\leq0.1\%$
for the sum and difference timestreams, respectively.
The sum and difference timestreams are polynomial-filtered and Fourier
transformed over each half-scan, and all auto-correlations and
cross-correlations between channel pairs are computed to form the
complex frequency-domain noise covariance matrix ${\bf { \tilde
N}}(f)$.  To form the noise model, we average ${\bf { \tilde N}}(f)$
over the 100 half-scans within each scan set and then average into 12
logarithmically spaced bins spanning 0.05--5~Hz.

To construct simulated correlated noise timestreams, we take the
Cholesky decomposition ${\bf { \tilde N}}(f) = {\bf L}(f){\bf L}^\dag
(f)$ of the noise covariance matrix and multiply a vector of
pseudo-random complex numbers $\boldsymbol{\rho}(f)$ by ${\bf L}(f)$:
\begin{equation}
{\bf {\tilde v}}(f) = {\bf L}(f)\boldsymbol{\rho}(f).
\end{equation}
The complex numbers in $\boldsymbol{\rho}(f)$ have Gaussian-distributed real and
imaginary components and are normalized such that the magnitude has a
standard deviation of one.  The resulting product ${\bf{\tilde v}}(f)$
has the same statistical properties as the data and is inverse Fourier
transformed to obtain a set of simulated noise timestreams.
Scan-synchronous templates are calculated and subtracted from each set
of azimuth scans, and the filtered noise timestreams are then co-added
into maps.

\begin{figure*}
\resizebox{\textwidth}{!}{\includegraphics{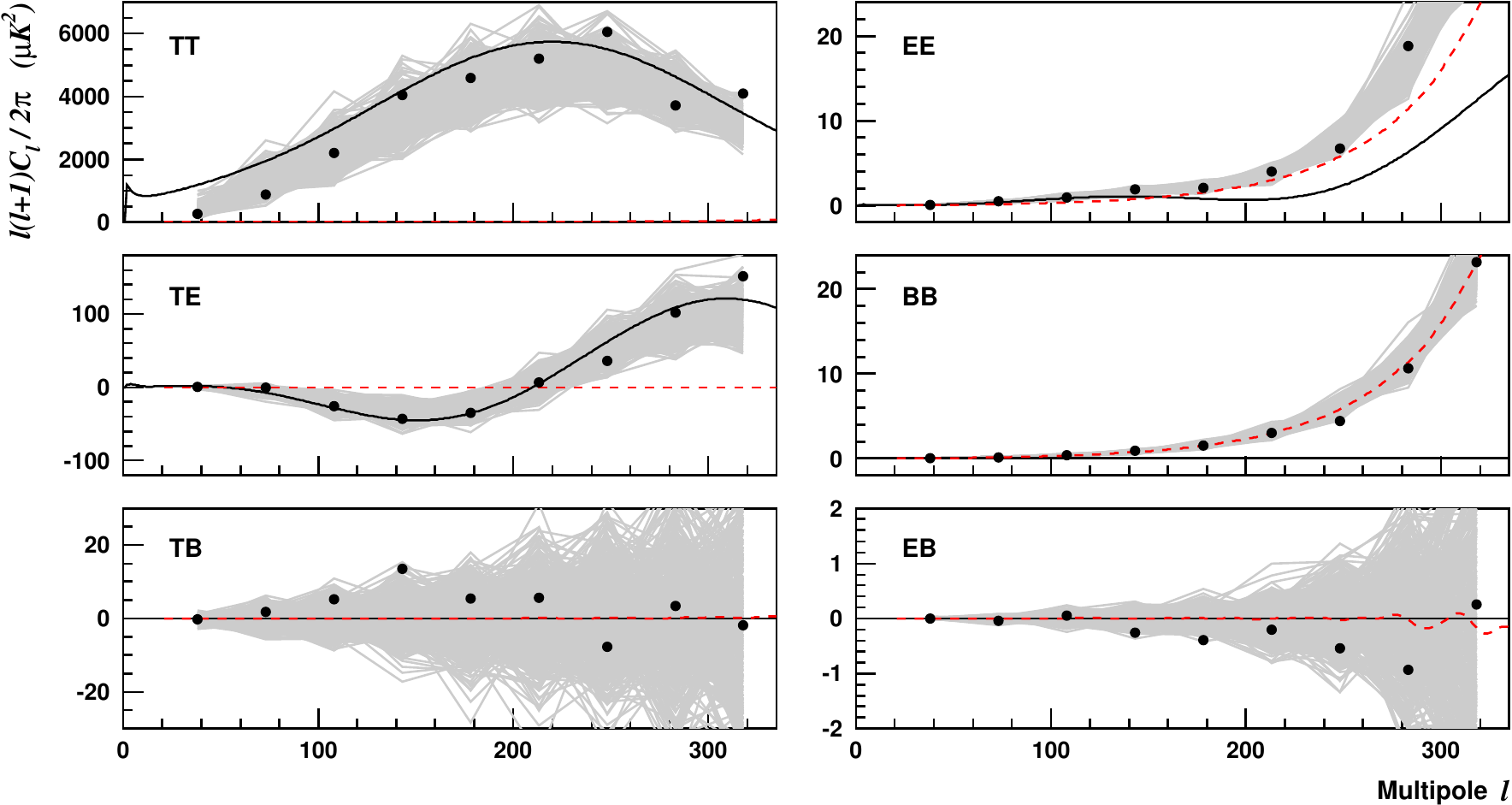}}
\caption{Raw power spectra, uncorrected for noise and filter bias, are
shown here for 500 signal-plus-noise simulations (gray lines) and
\bicep\ 150~GHz data (black points).  The dashed red lines indicate
the level of noise bias, $\ell(\ell+1){\hat N}_\ell/(2\pi)$, and the
black lines show the \lcdm\ spectra used as inputs for the signal
simulations.  Noise dominates much of the $EE$ spectrum and all of the
$BB$ spectrum.  The effect of filter bias is visible in the low-$\ell$
portion of the $TT$ spectrum, where the simulated spectra fall below
the model curve. \\}
\label{fig:sn_spectra}
\end{figure*}

The noise bias, ${\hat N}_\ell$, is estimated by averaging the power spectra
from an ensemble of noise-only maps.  Figure~\ref{fig:sn_spectra} shows ${\hat
  N}_\ell$ (red dashed lines), calculated from the average over 500
realizations, in comparison to raw power spectra.  \bicep\ measures $TT$ and
$TE$ with high S/N, and the noise contribution is negligible up to
$\ell\sim330$.  The noise is mostly uncorrelated between temperature and polarization,
so ${\hat N}_\ell$ for $TB$ is distributed around zero, and the same
is true for $EB$.  In contrast, noise comprises the bulk of the $EE$ and all of the $BB$
amplitude at $\ell>150$, which illustrates the need for careful noise modeling
and subtraction.  We have studied the impact of noise misestimation by intentionally
scaling the noise spectra~\citep{takahashi09}, and we find that the
largest possible error in ${\hat N}_\ell$ is $\pm 3\%$.

\subsection{Beam and Pixelization Corrections}\label{s:beamcorr}

The finite resolution of the telescope and pixelization of the maps
result in suppression of the observed power spectra at small angular
scales.  This suppression of power is described by ${\cal B}^2_\ell =
B^2_\ell H^2_\ell$, the product of the beam and pixel window
functions, and is illustrated in Figure~\ref{fig:fl}.  In this
analysis, we approximate $B_\ell$ as the Legendre transform of the
average beam within a given frequency band, which is assumed to be a
circular Gaussian.  The full widths at 100 and 150~GHz are 0.93$\deg$
and 0.60$\deg$, respectively.  The pixel window function $H_\ell$ is
supplied by \healpix\ and corresponds to nside=256.  In the case of \bicep, 
${\cal B}_\ell$ varies slowly with respect to the \spice\ kernel,
$\kappa^X_{\ell\ell'}$, and can therefore be pulled out of the convolution.
The observed power spectra, after noise subtraction, are divided by
${\cal B}_\ell^2$ to correct for the effects of beam suppression and
pixelization:
\begin{equation}
(\hat{C}_\ell^{X} - \hat{N}_\ell^X) / {\cal B}_\ell^2 \simeq
 \sum_{\ell'} \kappa^X_{\ell\ell'} F^X_{\ell'} C_{\ell'}^{X}.
\end{equation}

\subsection{Filter Corrections}\label{s:filtercorr}

The bolometer timestreams are cleaned by subtracting a third-order
polynomial and scan-synchronous template; this cleaning procedure also
has the effect of removing large-scale CMB signal.  The amount of
signal loss is described by $F_\ell$, the $\ell$-space transfer
function imposed by timestream filtering.  In addition, although the
\spice\ estimator is unbiased in the mean, the timestream processing
removes spatial modes from the observed $T$, $Q$, and $U$ maps,
generically introducing couplings between the observed $E$ and
$B$ spectra.  These couplings are small, but are of
interest for the leakage of the relatively large \emode\ signal into
\bmode\ polarization.  
We characterize both $F_\ell$ and $E$-to-$B$
leakage with signal-only simulations.

The signal simulation procedure begins by generating model power
spectra using the \camb~\citep{lewis00} software package.  We use
\lcdm\ parameters derived from \wmap\ five-year data~\citep{hinshaw09}
and a tensor-to-scalar ratio of zero.  From the model spectra, we use
the \synfast\ utility in \healpix\ to create an ensemble of simulated
CMB skies pixelized at 0.11$\deg$ resolution (nside=512) and convolved
with the \bicep\ beams.  Actual pointing data are used to generate
smoothly interpolated PSB timestreams from the simulated $T, Q, U$
maps and their derivatives.  A PSB timestream sample $d({\bf p})$ that
falls into a pixel centered at ${\bf p}$ is expressed as a convolution
of the beam ${\cal P}({\bf r}-{\bf r}_b)$, which is centered at ${\bf
r}_b$, with a second-order Taylor expansion of the sky signal $m$
around ${\bf p}$~\citep{hivonponthieu09}:
\begin{eqnarray}
\nonumber d({\bf p}) = && \int d{\bf r} \: {\cal P}({\bf r}-{\bf r}_b)
{\Big [} m({\bf p}) + \grad m({\bf p})({\bf r}-{\bf p}) \\
&& + {1\over2} ({\bf r}-{\bf p})^T D^2m({\bf p}) ({\bf r}-{\bf p}) {\Big ]}.
\label{eq:bsky_convolve}
\end{eqnarray}
Here, $m({\bf p}) = g [ T({\bf p}) + \gamma(Q({\bf p})\cos2\psi + U({\bf
p})\sin2\psi) ]$, and $\grad$ and $D^2$ denote the first and second
derivatives in spherical coordinates.
Assuming Gaussian beams for each frequency band, 
Equation~\ref{eq:bsky_convolve} reduces to
\begin{eqnarray}
\nonumber d({\bf p}) = && m({\bf p}) + \grad m({\bf p}) 
({\bf r}_b - {\bf p}) + \\
&& {1\over2} {\rm Tr} {\Bigg [} D^2 m({\bf p}) 
\left( \begin{array}{cc}
(\Delta \phi)^2 & \Delta \phi \Delta \theta \\
\Delta \phi \Delta \theta & (\Delta \theta)^2 \\
\end{array}\right )
{\Bigg ]},
\label{eq:bsky_gaussbeam}
\end{eqnarray}
where $\Delta\phi$ and $\Delta\theta$ are the longitude and latitude
offsets between the pointing vector ${\bf r}_b$ and the pixel center
${\bf p}$.  We apply Equation~\ref{eq:bsky_gaussbeam} to simulate
signal-only detector timestreams according to \bicep's scan strategy.
Measured PSB pair centroids, detector orientation angles, and
cross-polar leakage values are included in the simulations.  For the
purpose of characterizing the effects of timestream filtering, all
differential beam systematic effects are turned off so that there is
no mixing between temperature and polarization.  The simulated
timestreams are filtered and weighted in exactly the same way as the
real data and then co-added into maps.

Once we have ``\bicep-observed'' signal-only maps in hand, we compute
the power spectra with \spice\ and average the results over many
realizations.  Because the input \camb\ spectra have $r = 0$, any
non-zero $BB$ power in the simulation outputs is interpreted as
contamination from \emode\ polarization that is induced by timestream
filtering.  We therefore apply a correction,
\begin{equation}
(\hat{C}_\ell^{BB} - \hat{N}_\ell^{BB} -
 \hat{C}_\ell^{BB}\arrowvert_{EE~{\rm only}}) / {\cal B}_\ell^2 \simeq
 \sum_{\ell'} \kappa^{BB}_{\ell\ell'} F^{BB}_{\ell'} C_{\ell'}^{BB},
\label{eq:bb_debias}
\end{equation}
to the $BB$ power spectrum of the data, where
$\hat{C}_\ell^{BB}\arrowvert_{EE~{\rm only}}$ is the ensemble average
$BB$ spectrum from the $r = 0$ signal simulation outputs.  The
amplitude of this correction is roughly
$\ell(\ell+1)\hat{C}_\ell^{BB}\arrowvert_{EE~{\rm only}}/(2\pi) \simeq
3\times 10^{-3} \muK^2$ at $\ell\sim100$, comparable to inflationary 
$BB$ power for $r=0.05$,
and the sample variance is about $3\times 10^{-3} \muK^2$.
The correction factors for the other spectra
are negligibly small, so only the $BB$ spectrum is adjusted with this
procedure.

\begin{figure}[t]
\resizebox{\columnwidth}{!}{\includegraphics{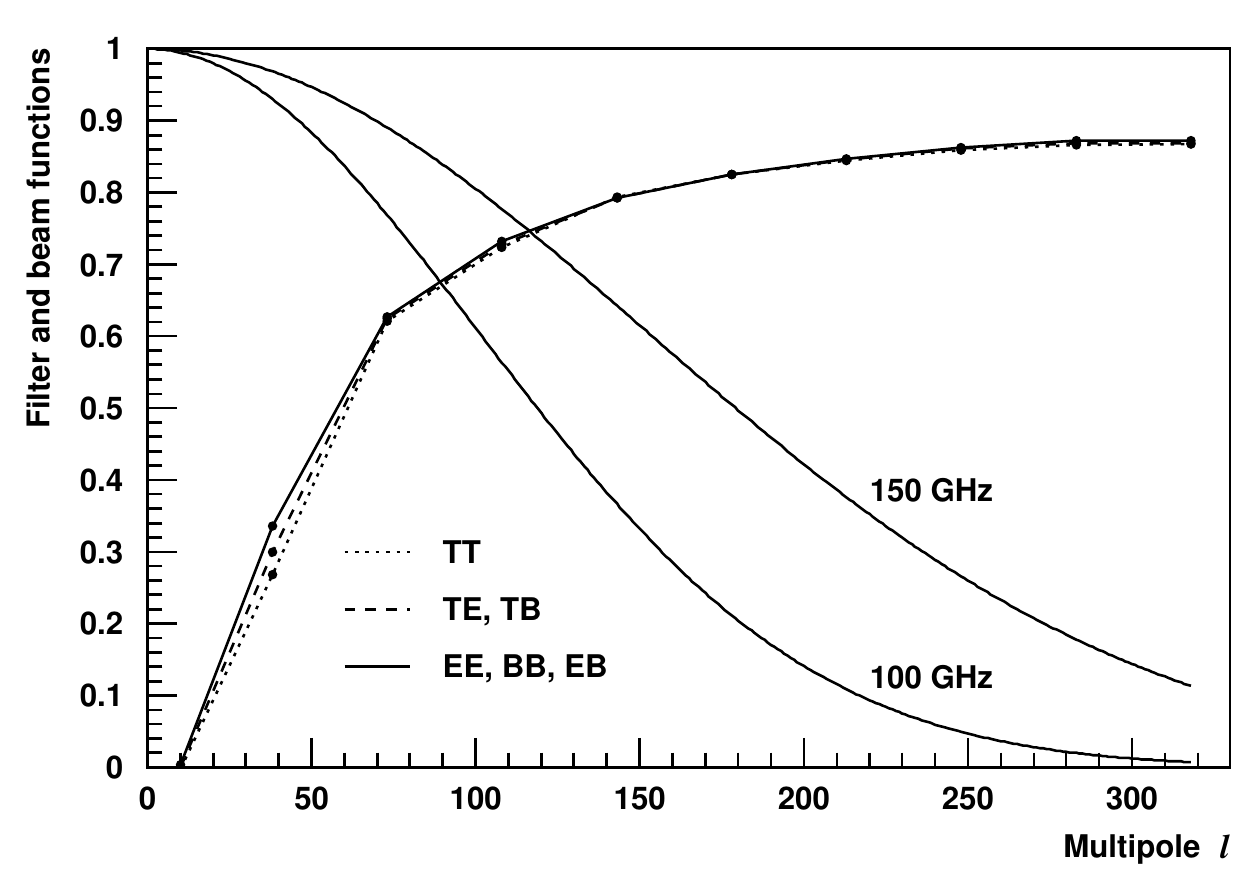}}
\caption{Signal-only simulations are used to evaluate the filter
function $F^X_b$, shown here for 150~GHz ($F^X_b$ for other frequency
combinations look similar).  The beam--pixel window functions ${\cal
B}_\ell^2 = B_\ell^2 H_\ell^2$ are shown for 100 and 150~GHz. \\}
\label{fig:fl}
\end{figure}

To quantify the suppression of large-scale power from timestream
filtering, we begin by assuming that $F^X_\ell$, like ${\cal
B}^2_\ell$, varies slowly in comparison to the \spice\ kernel and can
be pulled out of the convolution.  We choose to calculate the binned
filter suppression factor, $F^X_b$, defined as
\begin{equation}
\sum_{\ell} P_\ell^b \frac{\ell(\ell+1)}{2\pi}
\frac{(\hat{C}_\ell^{X} - \hat{N}_\ell^X)}{{\cal B}_\ell^2} \simeq 
F_b^X \sum_\ell P_\ell^b ~\frac{\ell(\ell+1)}{2\pi} \sum_{\ell'}
\kappa_{\ell\ell'} C_{\ell'}^X,
\label{eq:binspec}
\end{equation}
where the binning operator $P_\ell^b$ is a top hat.  (In the case of
the $BB$ spectrum, the left hand side of the equation also contains
$\hat{C}_\ell^{BB}\arrowvert_{EE~{\rm only}}$, as in
Equation~\ref{eq:bb_debias}.)  We obtain the filter suppression
factors by comparing the power spectra of the \bicep-observed
signal-only maps to those of the input \synfast\ maps; $F^X_b$ is the
ratio of the spectra, after multiplying by $\ell(\ell+1)/(2\pi)$ and
binning.

Figure~\ref{fig:fl} shows $F^X_b$ averaged over 500 signal simulations
using \bicep's timestream filtering.  In each simulation, signal-only
timestreams are generated from the full two years of pointing data.
At $\ell\sim100$, the value of $F^X_b$ is about 0.7 for all spectra
and rises slowly as $\ell$ increases.  Identical filter functions are
used for the $EE$ and $BB$ spectra, and the filter functions of the
cross-spectra are calculated as the geometric mean of those determined
from the auto-spectra.
The validity of this approach for obtaining the $BB$ and cross-spectra
filter functions has been confirmed with simulations over a limited
multipole range.
Dividing by $F^X_b$ is the final step in obtaining the \bicep\ band
power estimates:
\begin{eqnarray}
\hat{\mathscr C}^X_b &\equiv& \frac{1}{F^X_b} \sum_\ell P^b_\ell
\frac{\ell(\ell+1)}{2\pi} \frac{(\hat{C}_\ell^{X} -
\hat{N}_\ell^X)}{{\cal B}_\ell^2} ~ , ~ X \neq BB \label{eq:bp_est} \\
\hat{\mathscr C}^{BB}_b &\equiv& \frac{1}{F^{BB}_b} \sum_\ell P^b_\ell
\frac{\ell(\ell+1)}{2\pi} \frac{(\hat{C}_\ell^{BB} - \hat{N}_\ell^{BB} -
\hat{C}_\ell^{BB}\arrowvert_{EE~{\rm only}})}{{\cal B}_\ell^2}. \label{eq:bp_est_bb}
\end{eqnarray}

\subsection{Error Bars}\label{s:errors}

The uncertainties in the power spectra consist of two components, one that is
proportional to the signal itself (sample variance), and another that
depends on the instrumental noise.  We estimate the errors by
examining the variance of power spectra from simulated
signal-plus-noise maps, which exactly encode time-dependent correlated
noise, scan strategy, and sky coverage.  We add the simulated
noise-only and signal-only maps, described in \S\ref{s:nsim} and
\S\ref{s:filtercorr}; and power spectra are calculated for each
realization using the same ${\hat N}_\ell$, ${\cal B}_\ell^2$, and
$F^X_b$ as applied to the real data.  If the simulations include a
reasonable model of the signal and faithfully reproduce all the noise
properties of the experiment, then the data and simulations should be
indistinguishable.  Figure~\ref{fig:sn_spectra} shows raw power
spectra, uncorrected for noise and filter bias, from 500
signal-plus-noise simulations at 150~GHz.  The raw spectra of the
actual data, shown by the black points, lie within the scatter of the
simulations.  We calculate the band power covariance matrix from the
ensemble of simulations, after applying noise, filter, and beam
corrections.  The band power errors are obtained from square root of
the diagonal terms of the matrix.

\subsection{Band Power Window Functions}\label{s:bpwf}

\begin{figure}[t]
\resizebox{\columnwidth}{!}{\includegraphics{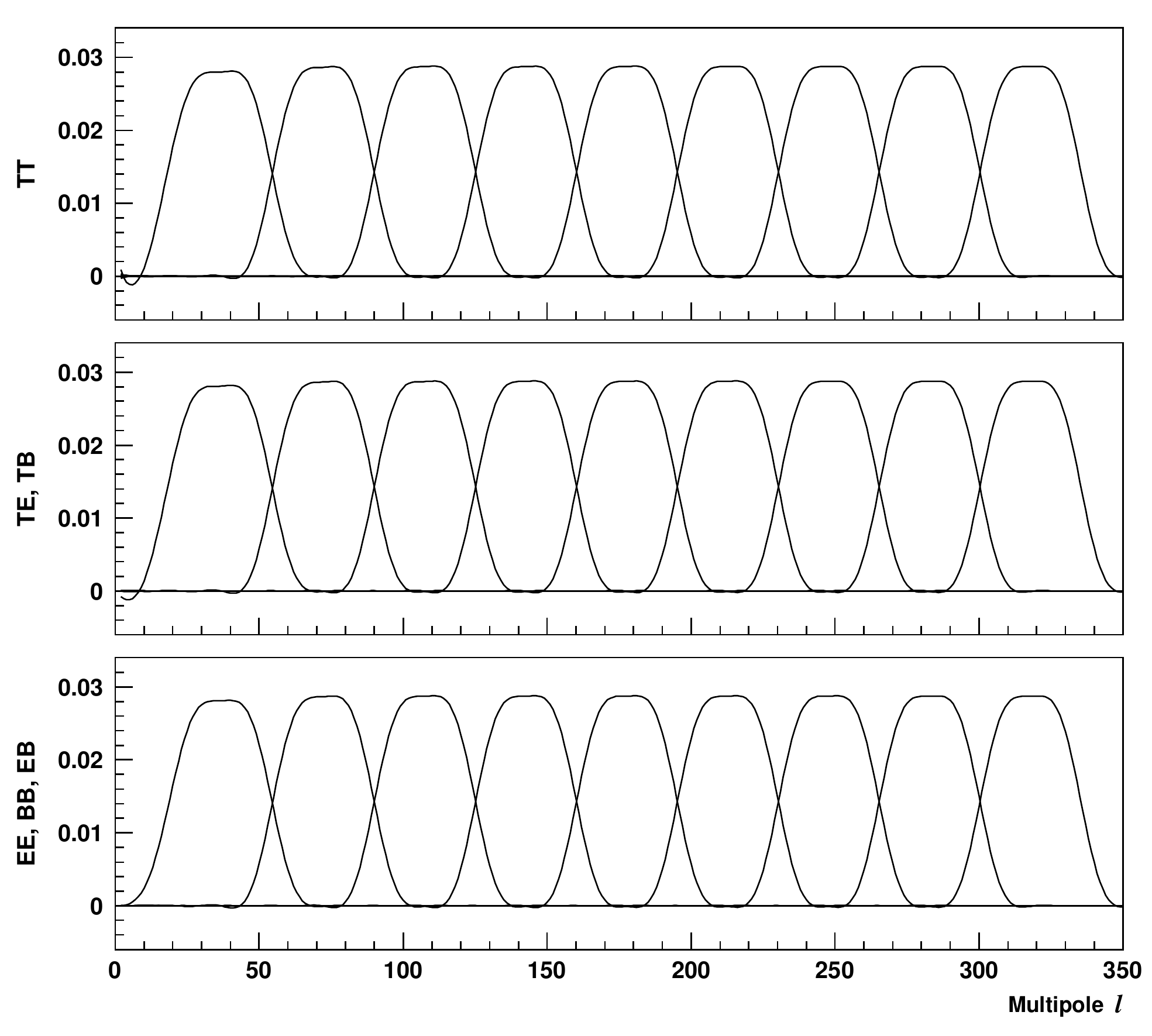}}
\caption{\bicep's band power window functions are defined by
Equation~\ref{eq:wdef}, and this figure shows $\omega_\ell^b /
\ell$. \\}
\label{fig:bpwf}
\end{figure}

In order to compare our band power estimates to a theoretical model,
we need a method to calculate expected band power values from 
theoretical power spectra.  The relationship between the model and the
expected band powers is described by band power window functions,
$\omega_\ell^b$, defined as
\begin{equation}
{\mathscr C}_b = \sum_\ell \frac{(\ell+\frac{1}{2})}{\ell (\ell+1)} ~\omega_\ell^b
~\mathscr{C}_\ell,
\label{eq:wdef}
\end{equation}
where $\mathscr{C}_\ell \equiv \ell (\ell+1) C_\ell / (2\pi) $.
The window functions are given by
\begin{equation}
\omega_\ell^b = \frac{2}{2\ell+1} \sum_{\ell'} P^b_{\ell'} \ell'
(\ell'+1) \kappa_{\ell'\ell}
\end{equation}
and are normalized such that
\begin{equation}
\sum_\ell \frac{(\ell+\frac{1}{2})}{\ell (\ell+1)} ~\omega_\ell^b = 1.
\end{equation}
The window functions depend on the \spice\ kernel, which depends on
the apodization function applied to the correlation functions.  For
this analysis, we apodize the correlation functions with a cosine
window that spans 50$\deg$.  We choose to use a uniform binning of
width $\Delta\ell = 35$, spanning a multipole range of $21 \leq \ell
\leq 335$.  This choice of bin width provides minimally correlated
band powers while preserving the spectral resolution determined by the
width of the \spice\ kernel.  Figure~\ref{fig:bpwf} illustrates
\bicep's band power window functions for the nine $\ell$~bins.

Because the $BB$ power spectrum is debiased with the procedure
described in Equation~\ref{eq:bb_debias}, we set the $EE$-to-$BB$
window functions to zero. This is a valid approximation as long as the $EE$
spectrum of the signal under consideration is statistically consistent with 
the measured \bicep\ band powers (which, as shown in \S\ref{s:datalcdm}, 
are well described by a concordance \lcdm\ cosmology).

\section{Power Spectrum Results}

\begin{figure*}[h]
\begin{center}
\resizebox{0.8\textwidth}{!}{\includegraphics{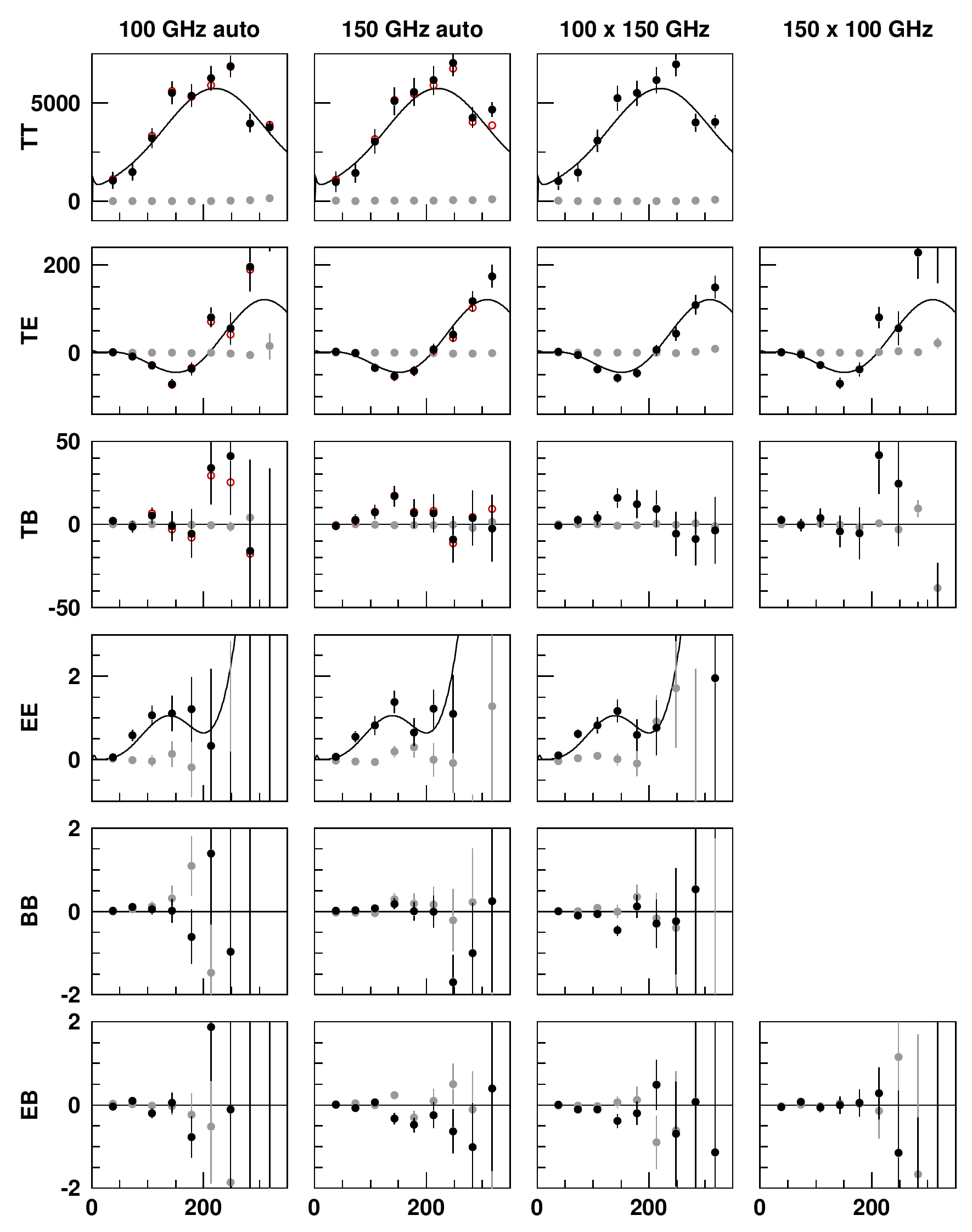}}
\caption{Black points show the full set of \bicep's power spectra.
The horizontal axis is multipole moment $\ell$, and the vertical axis
is $\ell(\ell+1)C_\ell/(2\pi)$ in units of$\muK^2$.  The spectra agree
well with a \lcdm\ model (black lines) derived from \wmap\ five-year 
data and $r = 0$.  The gray points correspond to the boresight
angle pair jackknife; note that although the $TT$, $TE$, and $TB$
jackknife failures are statistically significant, the amplitudes are
small compared to the signal.  The open circles show spectra
calculated using \wmap\ five-year temperature data, smoothed and
filtered identically to \bicep.  We have cross-correlated the \wmap\
Q and V-band data in \bicep's CMB field to obtain the $TT$ points,
and the $TE$ and $TB$ spectra are calculated from the cross-correlation
of \wmap\ V-band temperature data with \bicep\ polarization data.}
\label{fig:all_spectra}
\end{center}
\end{figure*}

Figure~\ref{fig:all_spectra} shows the full set of \bicep\ spectra
plotted with a \lcdm\ model derived from \wmap\ five-year data.  The
100 and 150-GHz auto-spectra are shown, as well as the 100$\times$150
cross-spectra.  In the case of the $TE$, $TB$, and $EB$ spectra, we
also show the 150$\times$100 cross-correlation.  For each spectrum, we
present nine band powers with a uniform bin width of $\Delta\ell =
35$, spanning $21 \leq \ell \leq 335$.  The $TT$, $TE$, and $EE$
spectra are detected with high significance and are already
sample-variance limited, and there is no detection of signal in $BB$,
$TB$, and $EB$.  The results from \bicep's two analysis pipelines
agree well with each other, and Figure~\ref{fig:data_vs_lcdm} shows a
comparison of the frequency-combined power
spectra~(\S\ref{s:combineps}).

As a cross-check, we have also derived $TT$, $TE$, and $TB$ spectra
using \wmap\ five-year temperature data in \bicep's CMB field (open circles
in Figure~\ref{fig:all_spectra}).  The \wmap\ temperature maps
are smoothed and filtered identically to \bicep, as described in
\S\ref{s:gains}.  The $TT$ points are calculated from the
cross-correlation of the \wmap\ Q and V-band maps, and the $TE$ and
$TB$ spectra are calculated using \wmap\ V-band temperature data and
\bicep\ polarization data.  For this comparison, we do not subtract
noise bias and instead rely on the fact that the pairs of maps have
uncorrelated noise.  We also do not attempt to assign error bars.
Qualitatively, the spectra formed using \wmap\ temperature data agree
well with the spectra from \bicep\ temperature and polarization data.
Both the \bicep\ and \wmap\ temperature maps are strongly signal-dominated;
apparently the differences between them, including the noise as well as
potential systematics, are at a level that has little impact on 
the $TT$, $TE$, or $TB$ power spectra results.

\section{Consistency Tests}

\subsection{Jackknife Descriptions}\label{s:jacktypes}

We check the self-consistency of the power spectra by performing
jackknives, statistical tests in which the data are split in two
halves and differenced.  The split is performed at the mapmaking
stage, and the resulting differenced map should have power spectra
that are consistent with the expected residual signal level (nearly
zero) after subtracting noise bias.  The interaction of timestream
filtering with the details of the split causes imperfect signal
cancellation when forming jackknife maps, but in practice, this
residual signal is small.

The data are tested with six jackknives that are sensitive to
different aspects of the instrument's performance.  In the scan
direction jackknife, the data are split into left- and right-going
azimuth half-scans.  Failures generally point to a problem in the
detector transfer function deconvolution, or thermal instabilities
created at the scan endpoints.  The elevation coverage jackknife is
formed from the two CMB observations in each 48-hr cycle; each
observation covers the same azimuth range but starts from a different
elevation.  This jackknife is sensitive to ground-fixed or
scan-synchronous contamination.  \bicep\ observes at four fixed
boresight orientation angles that can be split into two pairs,
$\{-45\deg, 0\deg\}$ and $\{135\deg, 180\deg\}$, to form a boresight
angle pair jackknife.  This test is perhaps the most powerful of the
jackknives performed and is sensitive to many factors, including
thermal stability, atmospheric opacity, relative gain mismatches,
differential beam pointing, and ground pickup.  In the temporal
jackknife, the 8-day observing cycles---48~hr at each of the four
boresight angles---are interleaved to form the two halves, and this
jackknife tests sensitivity to weather changes.  The season split
jackknife simply divides the data into the two observing seasons, and
failures reflect any changes made to the instrument between the two
years.  In particular, the focal plane thermal architecture was
improved for the 2007 season, and the temperature control scheme was
changed.  The focal plane $QU$ jackknife splits the detectors into two
groups according to their polarization orientation within the focal
plane (approximately alternating hextants) and is a method of probing
instrumental polarization effects.

Power spectra of jackknife maps are computed with the method described
in \S\ref{s:ps_estimation}, using simulated jackknife noise and
signal-plus-noise maps to subtract noise bias and assign error bars.
The filter function $F^X_b$ is applied so that the magnitude of any
non-zero jackknife band powers can be compared to the amplitude of the
non-jackknife spectra.  For each jackknife spectrum, we calculate
$\chi^2$ over nine bins spanning $21 \leq \ell \leq 335$.  To account
for the expected level of residual signal, the $\chi^2$ values are
evaluated with respect to the average jackknife spectra from an
ensemble of signal-only simulations.  The criteria for jackknife
success or failure are based on the probability to exceed (PTE) the
$\chi^2$ value, which is calculated from the distribution of $\chi^2$
in 500 signal-plus-noise simulations.  Jackknife victory is declared
when (1) none of the PTEs are abnormally high or low, given the number
of $\chi^2$ tests performed, and (2) the PTEs are consistent with a
uniform distribution between zero and one.

\subsection{Jackknife Results}

\begin{figure}[t]
\resizebox{\columnwidth}{!}{\includegraphics{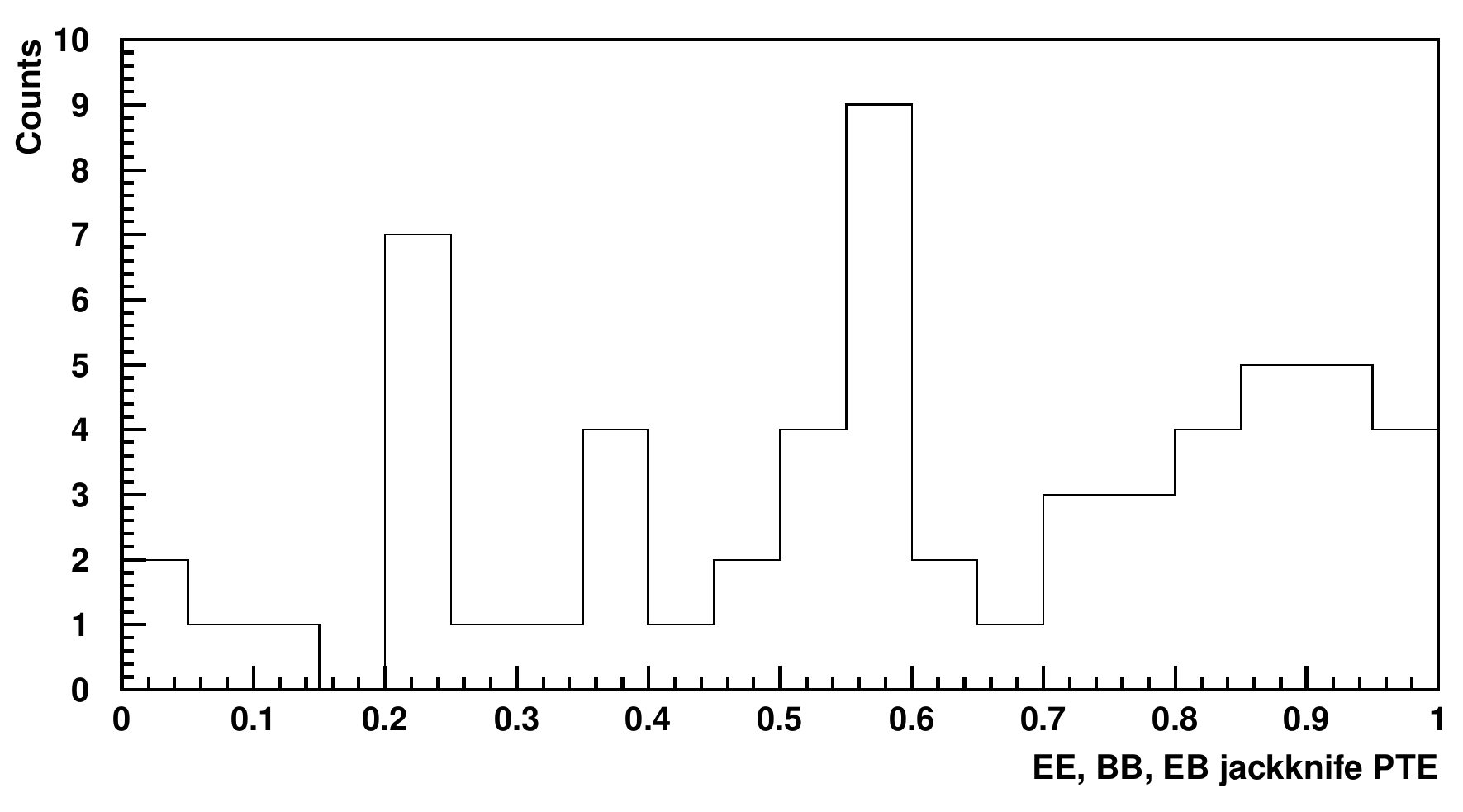}}
\caption{Probabilities to exceed the $\chi^2$ values from $EE$,
$BB$, and $EB$ jackknife tests are consistent with a uniform
distribution between zero and one. \\}
\label{fig:pte_dist}
\end{figure}

The jackknife spectra for the six different data splits appear similar
in that the band powers are all distributed around zero.  We therefore
show only the spectra from the boresight pair jackknife
(Figure~\ref{fig:all_spectra}, gray points), which is arguably the
most stringent of the six tests.  Table~\ref{t:ptes} lists the PTE
values from all of the jackknife $\chi^2$ tests for \bicep's
polarization-only spectra ($EE, BB, EB$).  Out of all the PTEs, the
lowest value is 0.014, in the 150~GHz focal plane $QU$ jackknife.
This low value is, however, consistent with expectations from
uniformly distributed PTEs over the 60 polarization-only jackknives.
Figure~\ref{fig:pte_dist} shows that the PTEs are consistent with a
uniform distribution between zero and one.

The polarization jackknife tests are the most powerful probe of the accuracy
of the noise model.  In addition to the $\chi^2$ tests, we have also expressed
the jackknife spectra as band power deviations with respect to the mean of
signal-only simulations.  The sum of the band power deviations in each
jackknife spectrum provides an additional and more precise gauge of the
correct estimation of instrument noise.  We have verified that the band power
deviation sums in the data are consistent with signal-plus-noise simulations
and are not systematically biased high or low, thus confirming that the noise
levels are correctly estimated.  Furthermore, we have
probed the sensitivity to systematic variations in the modeled
noise amplitude over a range that is comparable to the S/N of the sum and
difference timestreams, and find no significant changes in the jackknife
spectra.   The relative insensitivity of the jackknifes to the amplitude 
of the noise model validates the procedure described in \S\ref{s:nsim}.

In contrast to the polarization data, the temperature data display
significant jackknife failures.  There is an excess of small PTE
values in the $TE$ and $TB$ jackknives, and most of the $TT$ jackknife
PTEs are smaller than 0.002, which is the resolution from 500
simulations.  The $TT$, $TE$, and $TB$ jackknife PTEs are therefore
not listed in Table~\ref{t:ptes}.  We attribute these failures to the
fact that \bicep's temperature maps have high S/N (see
$TT$ plot in Figure~\ref{fig:sn_spectra}), and the jackknives are
therefore extremely sensitive to small gain calibration errors or
imperfections in modeling and subtracting unpolarized atmospheric
emission.  (As described in \S\ref{s:nsim}, the $\leq10\%$ S/N in the
pair-sum timestreams is a known imperfection of the \bicep\ noise
model for temperature data.)  Although the $TT$, $TE$, and $TB$
jackknife failures are statistically significant,
Figure~\ref{fig:all_spectra} illustrates that the amplitudes of the
jackknife spectra are small compared to both the amplitude and errors
of the signal spectra.  The magnitudes of the $TT$ and $TE$/$TB$
jackknife band powers are typically 1--10$\muK^2$ and 0.1--1$\muK^2$,
respectively.  In all cases, the error bars of the non-jackknife
spectra are at least a factor of 10 larger, except in the highest
$\ell$ bin.

We have performed the same jackknives with both analysis pipelines,
and the results are in excellent agreement.  The alternate pipeline
confirms that the polarization data pass the $\chi^2$ tests, and
although the $TT$, $TE$, and $TB$ spectra do not pass, the amplitudes
of the jackknife spectra are small.

\begin{deluxetable}{c c c c c}
\tablecaption{Jackknife PTE Values from $\chi^2$ Tests \label{t:ptes}}
\tablehead{\colhead{Jackknife} &
\colhead{100~GHz} & \colhead{150~GHz} & \colhead{100$\times$150} & \colhead{150$\times$100}}
\startdata
\input{ptetab}
\enddata
\end{deluxetable}

\section{Systematic Uncertainties}\label{s:systematics}

\begin{figure}[t]
\resizebox{\columnwidth}{!}{\includegraphics{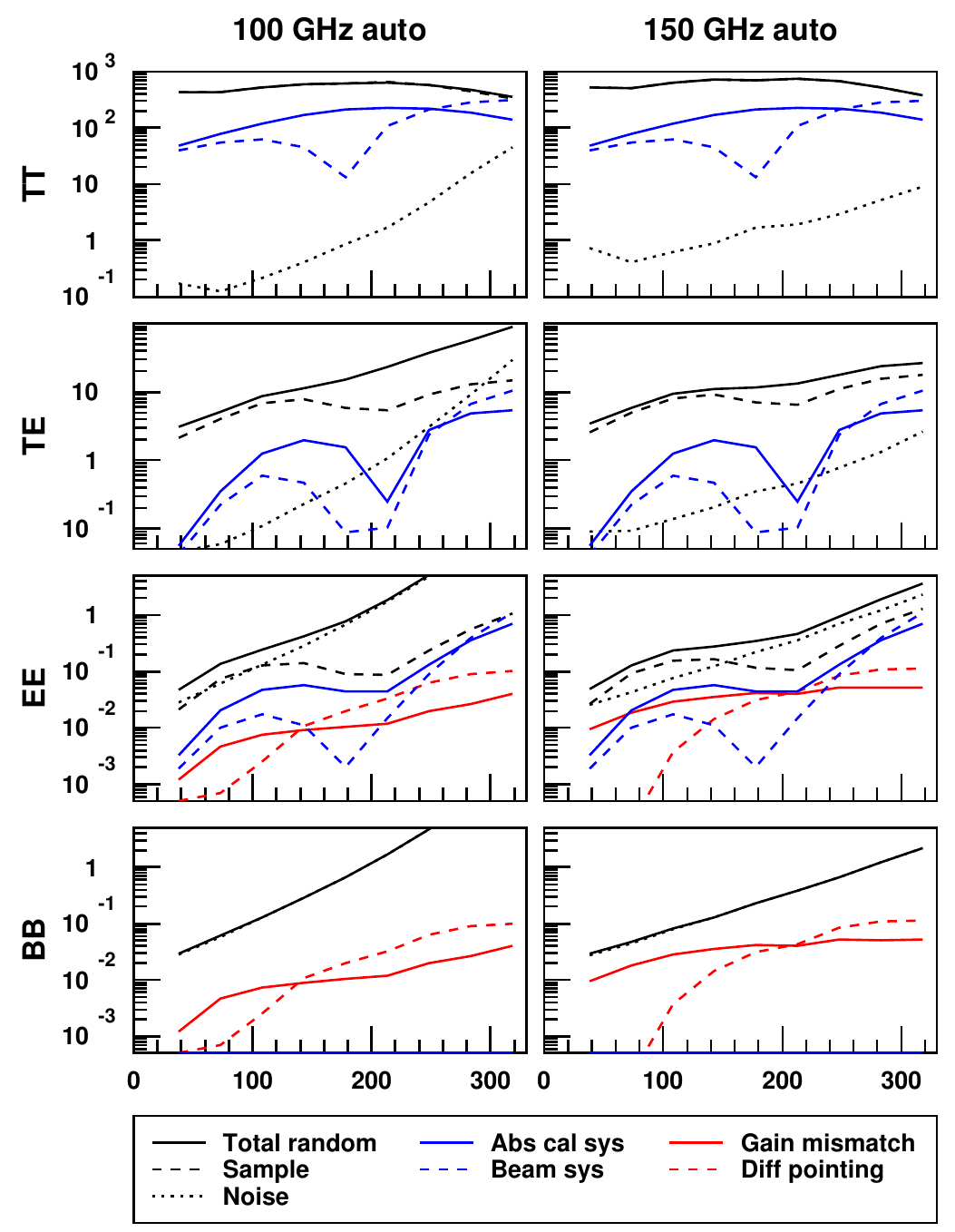}}
\caption{Dominant systematic uncertainties in \bicep's band powers
are compared to the total statistical error, which comes from sample
variance and noise.  The horizontal axis is multipole moment $\ell$,
and the vertical axis is $\ell(\ell+1)C_\ell/(2\pi)$ uncertainty in
units of$\muK^2$.  The contributions from the signal--noise
cross terms to the total statistical error in $TE$ are not 
shown.  We have assumed that gain mismatch and differential pointing
systematics leak temperature to $EE$ and $BB$ in equal amounts. \\}
\label{fig:errbudget}
\end{figure}

Systematic errors that arise from uncertainties in instrument
characterization can be separated into two categories: (1) errors that
mix temperature, \emode, and \bmode\ polarization; and (2) calibration
uncertainties that affect only the scaling or amplitude of the
spectra.  In this section, we describe the dominant sources of
systematic error in \bicep\ and the expected impact on the power
spectra; these systematics are summarized in
Figure~\ref{fig:errbudget}.  A complete description of all potential
\bicep\ systematics and the methodology for propagating the errors to
the power spectra is given in the accompanying instrument
characterization paper~\citep{takahashi09}.

\subsection{Temperature and Polarization Mixing}

We are mostly concerned with systematic errors that mix the bright
temperature signal into polarization and thus induce a false \bmode\
signal.  We define a benchmark for each systematic error such that the
false \bmode\ amplitude is no greater than the peak of the
inflationary $BB$ spectrum.  For \bicep's target of $r=0.1$, this
requirement corresponds to $\ell(\ell+1)C^{BB}_\ell/(2\pi) = 7 \times
10^{-3}\muK^2$ at $\ell\sim100$.

The primary source of potential temperature leakage into polarization
is differential gains within PSB pairs.  Gain mismatch effects can
also arise from other systematics, such as errors in the bolometer
transfer functions, which act as frequency-dependent gains.  We have
characterized the common-mode rejection of PSBs by examining the
cross-correlation of pair-sum and pair-difference maps.  There is no
statistically significant evidence for gain mismatch in the data, and
we set an upper limit of $\Delta(g_1/g_2)/(g_1/g_2) < 1.1\%$ on
differential gains.

A second source of potential temperature and polarization mixing is
beam mismatch within PSB pairs.  We describe beam mismatch with three
quantities: differential beam size, pointing, and ellipticity.  Of
these three, the dominant effect in \bicep\ is differential pointing,
which is stable over time and has been measured with a median
amplitude of $({\bf r}_1-{\bf r}_2)/\bar{\sigma} = 1.3 \pm 0.4\%$,
where $\bar{\sigma}$ is the average Gaussian beam size within a pair
of PSBs.
Measured upper limits on differential size and ellipticity are negligible.

Most systematic errors interact with the scan strategy in complex
ways, and the exact effects on the power spectra can be computed only
through signal simulations.  We follow the formalism presented in
\citet{hivonponthieu09} and \citet{shimon08} to calculate the expected
level of false $BB$ in \bicep.  Starting from \synfast\ maps with
$r=0$, Equation~\ref{eq:bsky_gaussbeam} is used to generate simulated
signal timestreams that include the effects of gain mismatch and
differential pointing.  The timestreams are filtered and co-added into
maps, and the amplitude of the $BB$ power spectrum at $\ell\sim100$ is
compared with the $7 \times 10^{-3}\muK^2$ benchmark.

In the differential gain simulations, we randomly assign 1.1\% rms
gain mismatch to PSB pairs, and we find that the expected false $BB$
amplitude is $7.4 \times 10^{-3}\muK^2$ and $2.9 \times 10^{-2}\muK^2$
at $\ell\sim100$ for 100 and 150~GHz, respectively.  Although this
amplitude exceeds the $r=0.1$ benchmark, Figure~\ref{fig:errbudget}
illustrates that it is small compared to the statistical error in this
analysis of the two-year data.  In a future analysis of the entire
\bicep\ data set, we anticipate placing tighter constraints on PSB
gain mismatch as the noise levels integrate down.  With further work,
we are confident that uncertainty in gain mismatch can be
substantially reduced from the current 1.1\%.

Differential pointing has been precisely characterized for each of
\bicep's PSB pairs, so we run signal simulations using the measured
centroid offset vectors, rather than randomized distributions.  The
false $BB$ from differential pointing has an $\ell\sim100$ amplitude
of $2.7\times10^{-3}\muK^2$ and $4.2\times10^{-3}\muK^2$ at 100 and
150~GHz, respectively.  These amplitudes are slightly smaller than the
$r=0.1$ benchmark and are well below the noise level of the initial
two-year data set.  In a future analysis, it may be possible to use
the measured centroid offsets to correct for systematic effects.

We emphasize that, in addition to the differential gain and pointing
discussed here, most uncertainties in instrument characterization {\it
create} false positive $BB$ signal.  The fact that \bicep's $BB$
spectra are consistent with zero and pass jackknives demonstrates that
we have achieved sufficient control over systematic errors in this
analysis.  Furthermore, until a positive \bmode\ detection is made,
the presence of systematic effects that produce spurious polarization 
could only make the reported $BB$ upper limits higher (more conservative)
than they would be otherwise.

\subsection{Absolute Gain and Beam Uncertainty}\label{s:gain_beam_sys}

The scaling of the power spectra is determined by the absolute gain
factors that convert detector units to temperature, and the 2\%
uncertainty in this gain (\S\ref{s:gains}) translates into a 4\%
uncertainty in the power spectrum amplitude.  The polarized spectra
have additional amplitude uncertainty that arises from errors in the
cross-polar leakage.  A systematic error of 
$\Delta\epsilon = 0.01$ corresponds to 3.9\% amplitude uncertainty for
polarization-only spectra ($EE$, $BB$, $EB$) and 2.0\% for
temperature--polarization cross-spectra ($TE$, $TB$).  Therefore, the
combined amplitude uncertainty $G^X \equiv \Delta C^X_\ell / C^X_\ell$
from gain and $\epsilon$ errors is 4\% for $TT$, 4.5\% for $TE$/$TB$,
and 5.6\% for $EE$/$BB$/$EB$.  These uncertainties are shown in
Figure~\ref{fig:errbudget}, assuming \lcdm\ spectra with $r=0$.

For Gaussian beams, measurement errors in the beam widths introduce a
fractional uncertainty,
\begin{equation}
S_\ell \equiv \Delta C^X_\ell / C^X_\ell = e^{\sigma^2 \ell
(\ell+1)(\delta^2+2\delta)} - 1 - {\bar S},
\label{eq:beam_err}
\end{equation}
in the power spectrum amplitude as a function of $\ell$.  Here,
$\sigma = {\rm FWHM} / \sqrt{8\ln(2)}$, and $\delta$ is the fractional beam
width error $\Delta\sigma/\sigma$.  Because this band power
uncertainty is degenerate with absolute gain error, we subtract the
mean, ${\bar S}$, calculated over $56 \le \ell \le 265$, the angular
scales over which we perform absolute calibration.  For this analysis,
we use average beam widths of 0.93$\deg$ and 0.60$\deg$ at 100 and
150~GHz, respectively.  Although the distribution of beam widths in
the focal plane varies by $\pm3\%$, the measurement precision is
$\pm0.5\%$; we therefore expect the effective $\delta$ to lie somewhere
in between.  We calculate the maximum beam width error allowed by our
calibration cross-spectra (Equation~\ref{eq:gain_cal}), which are very
close to flat, and we constrain $\delta < 1.2\%$ and $\delta < 2.8\%$ at 100 and
150~GHz, respectively.  Figure~\ref{fig:errbudget} shows the expected
power spectrum errors from these maximum allowed beam uncertainties,
which are most likely conservative.  The systematic error is smaller
than the statistical error in all cases except the $TT$ spectra at
high~$\ell$, where the levels of statistical and beam systematic
uncertainty are comparable.

\section{Foregrounds}\label{s:foregrounds}

The \bicep\ CMB region was chosen to have the lowest foreground dust
emission for a field of that size, and we do not expect foreground
contamination from dust or other sources to be significant at the
current depth in the maps.  To verify this, we estimate the levels of
contamination in \bicep\ CMB data from three potential foreground
sources: thermal dust emission, synchrotron radiation, and
extragalactic radio point sources.

\subsection{Thermal Dust}

\begin{figure}[t]
\resizebox{\columnwidth}{!}{\includegraphics{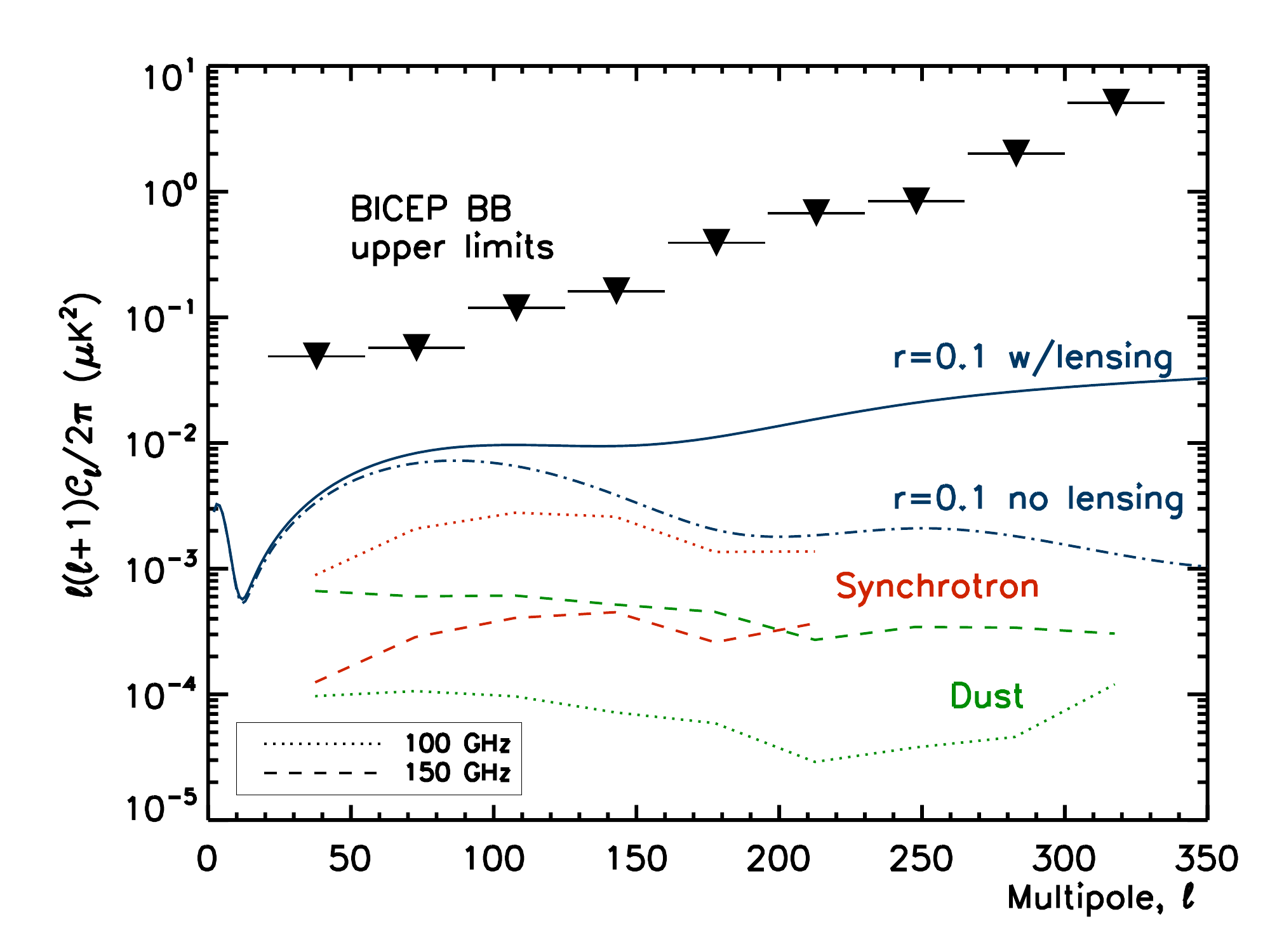}}
\caption{Expected levels of polarized dust and polarized synchrotron
  in the \bicep\ CMB field, assuming FDS Model~8 with $5\%$
  polarization fraction and \wmap\ MCMC polarization maps extrapolated
  to 100 and 150~GHz. These estimated
  foreground levels are much lower than the $BB$ upper limits
  presented in~\S\ref{s:t2s}.  Theoretical $BB$ curves for $r=0.1$, with and without 
  gravitational lensing, are also shown for comparison. \\}
\label{fig:foreground}
\end{figure}

Polarized dust emission in the \bicep\ CMB field is estimated from
``FDS Model~8''~\citep{finkbeiner99}.  We assume a 5\% polarized
fraction, guided by a study of \wmap\ data that shows that
high-latitude dust has a mean fractional polarization of
3.6\%~\citep{kogut07}.  The dust temperature and polarization model is
extrapolated to 100 and 150~GHz, and filtered according to \bicep's
scan strategy.  The resulting polarized dust emission is
$\ell(\ell+1)C_\ell/(2\pi) = 9.6\times10^{-5}\muK^2$ and
$6.1\times10^{-4}\muK^2$ at $\ell=100$ for 100 and 150~GHz,
respectively.  These values are 2 orders of magnitude below the 95\%
confidence level for upper limits on the $BB$ amplitude from \bicep\ 
(discussed in \S\ref{s:t2s}),
as shown in Figure~\ref{fig:foreground}.

We have also tested for dust contamination in the \bicep\ maps by
studying the cross power spectrum between FDS Model~8 maps and
\bicep\ temperature and polarization maps. The cross power spectrum of
the real data is consistent with the
distribution from signal-plus-noise simulations, providing 
additional evidence that the  \bicep\ maps contain no spatial
correlation with the FDS dust maps.

\subsection{Synchrotron}

To estimate the polarized synchrotron emission in the \bicep\ field of
view, we have used the \wmap\ Markov Chain Monte Carlo (MCMC)
synchrotron maps~\citep{gold08}, extrapolated to 100 and 150~GHz using
the mean spectral index of the \bicep\ field of view, calculated from the spectral index maps provided in the same analysis. As for the FDS
dust maps, we filter the extrapolated 100 and 150~GHz synchrotron maps
with the \bicep\ scan strategy, and find the estimated level of
polarized synchrotron emission at $\ell\sim100$ to be
$3\times10^{-3}\muK^2$ and $4\times10^{-4}\muK^2$ for 100 and 150~GHz,
respectively, both below the level of \bicep\ sensitivity. 
Furthermore, the \wmap\ MCMC map has poor S/N in regions far
from the Galactic plane, and we have found that within the \bicep\
CMB field, the map is dominated by variance in the Monte Carlo
fit \citep[see][for details]{gold08}.  The estimated levels of
polarized synchrotron emission in Figure~\ref{fig:foreground} should
therefore be viewed as conservative estimates or upper limits.
We have also derived synchrotron estimates
from \wmap\ K-band polarization maps and the temperature maps of
\citet{haslam81}, assuming a 30\% polarization fraction; both methods
give synchrotron estimates that are lower than those from the \wmap\
MCMC maps.

Similarly to our analysis for thermal dust, we have studied the cross
power spectrum of the synchrotron maps with the \bicep\ maps and found
there to be no significant spatial correlation of the \bicep\ data
with the synchrotron emission.
 
\subsection{Point Sources}

At degree-scale resolution, the \bicep\ maps do not show any obvious
point source detections, so we rely on a combination of the 4.85-GHz
Parkes--MIT--NRAO (PMN) survey~\citep{wright94}, the \wmap\ point
source catalog~\citep{wright09}, and the \acbar\
catalog~\citep{reichardt09} to search for point source
contamination. We search for point source contamination by optimally
filtering CMB and noise fluctuations out of the \bicep\ temperature
map and determine the significance of the resulting pixel values by
repeating the process with simulated maps of CMB with detector noise.
Although the resulting maps have a few 2$\sigma$ detections at the
suspected point source locations, there is no statistical evidence for
point source contamination above the expected Gaussian distribution of
noise.  As a further test, we have simulated the effects of masking
out the 27 \acbar\ sources that lie within the \bicep\ field and have
found it has no significant impact on the power spectra.

\subsection{Frequency Jackknife}

The CMB and foreground emission have different frequency dependence,
so we can test for the presence of foreground contamination in \bicep\
data by performing a frequency jackknife.  We difference the 100 and
150-GHz maps, compute the power spectra, calculate $\chi^2$, and
compare the results to signal-plus-noise simulations, as described in
\S\ref{s:jacktypes}.  The probabilities to exceed the $\chi^2$ values
are $\{0.050$, $0.152$, $0.732\}$ for $EE$, $BB$, and $EB$,
respectively.  We find no evidence for foreground contamination in the
frequency jackknives.

\section{Combined Spectra}\label{s:combineps}

We combine the spectra from the different observing frequencies by
taking a weighted average for each band power.  To obtain the weights,
we use signal-plus-noise simulations to calculate the covariance
matrices from the various frequency combinations (100, 150,
100$\times$150, and 150$\times$100~GHz).  There are three unique
combinations for $TT$, $EE$, and $BB$, and four combinations for the
other spectra.  The weights are calculated from the row/column sums of
the inverse of the covariance matrices.  The error bars of the
combined spectra are determined by applying the same combination
weights to signal-plus-noise simulations.  For fully noise-dominated
spectra (such as $BB$), the error bars of the combined spectra improve
by 10\%--40\% compared to the errors from 150~GHz data alone.

\begin{figure*}
\begin{center}
\resizebox{\textwidth}{!}{\includegraphics{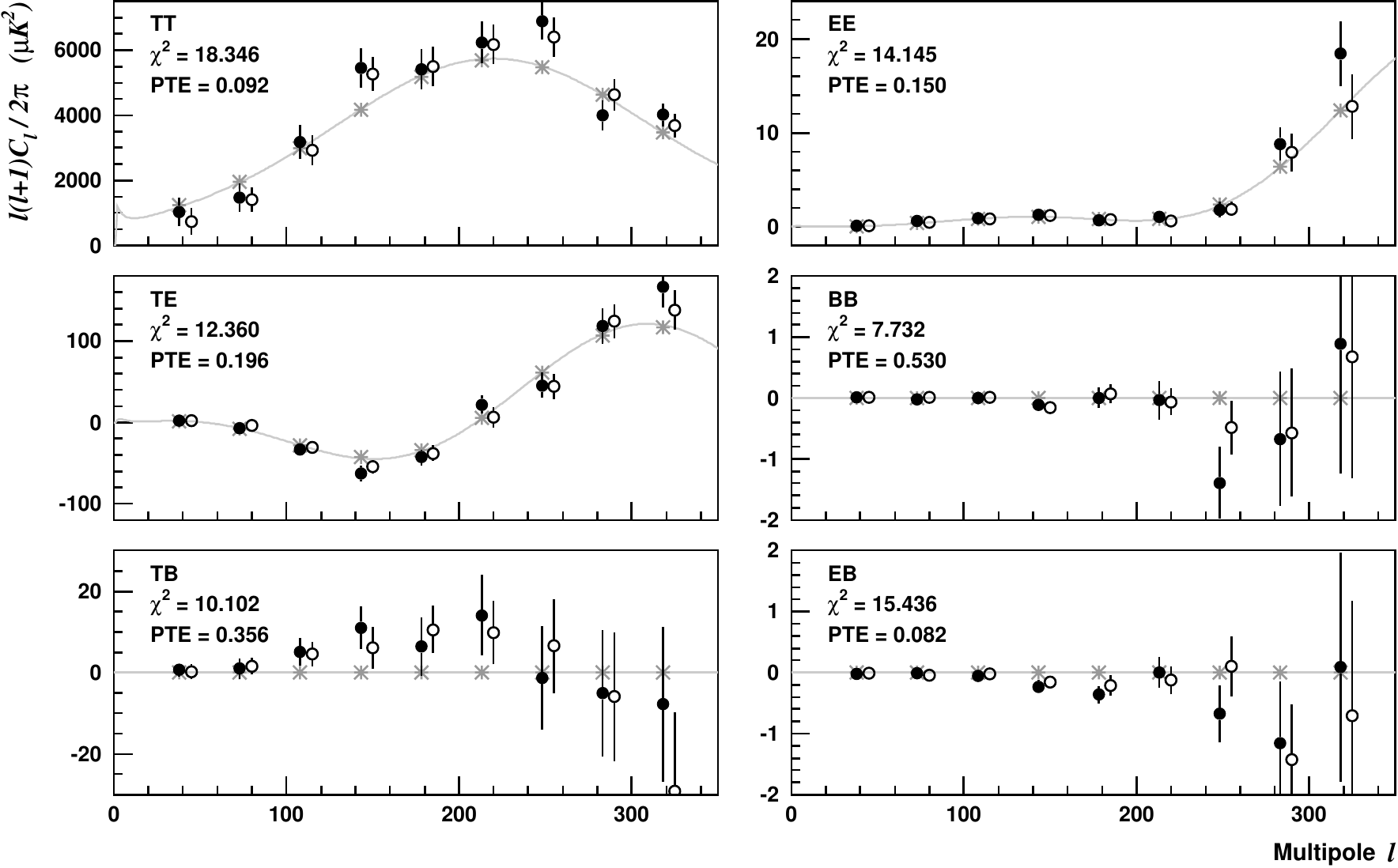}}
\caption{\bicep's combined power spectra (black points) are in
excellent agreement with a \lcdm\ model (gray lines) derived from
\wmap\ five-year data.  The $\chi^2$ (for nine degrees of
freedom) and PTE values from a comparison of the data with the model
are listed in the plots.  The asterisks denote theoretical
band power expectation values.
Power spectrum results from the alternate analysis
pipeline are shown by the open circles and are offset in $\ell$ for
clarity. \\}
\label{fig:data_vs_lcdm}
\end{center}
\end{figure*}

\begin{figure}[t]
\resizebox{\columnwidth}{!}{\includegraphics{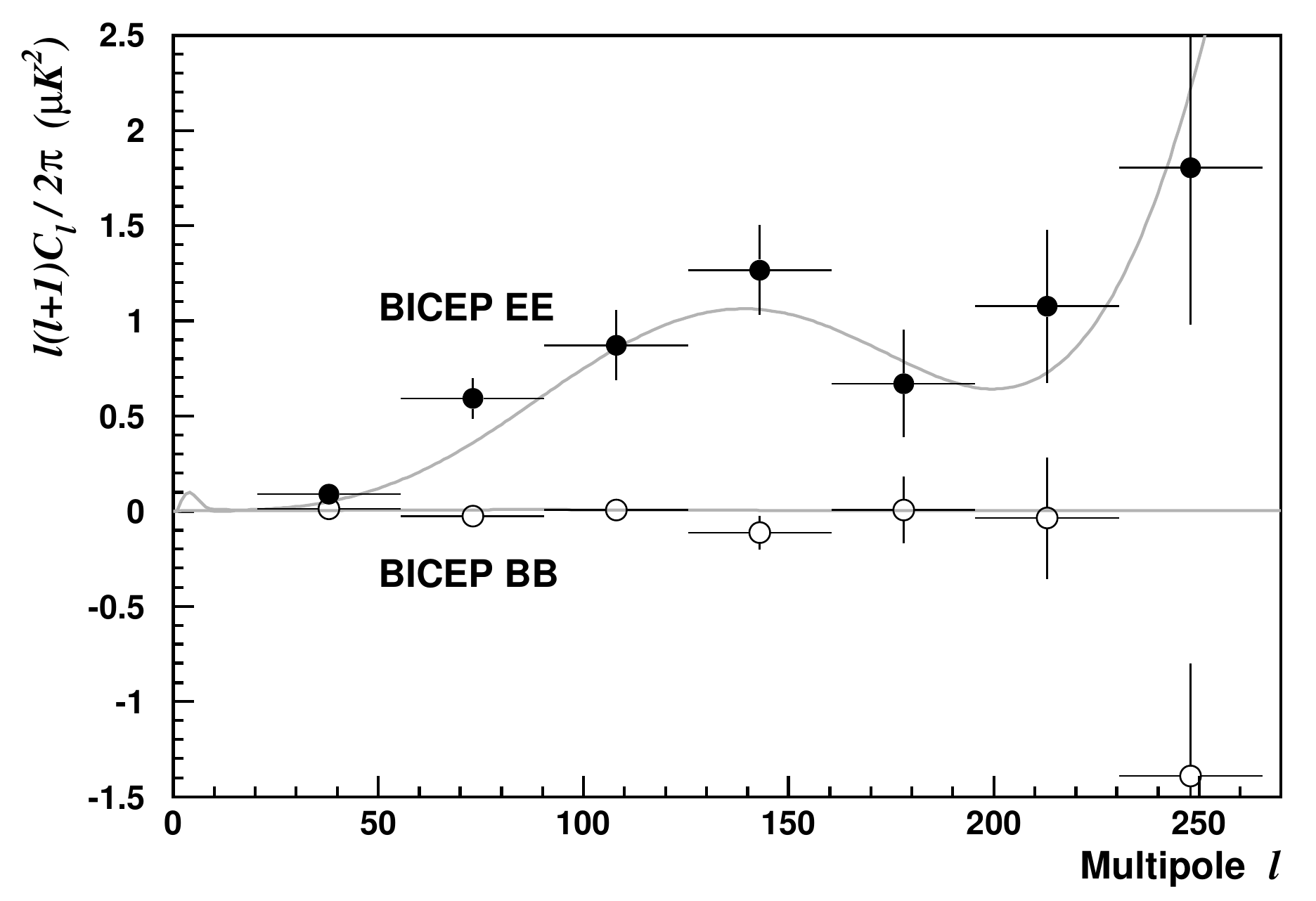}}
\caption{\bicep\ measures $EE$ polarization (black points) with high
S/N at degree angular scales.  The $BB$ spectrum (open
circles) is overplotted and is consistent with zero.  Theoretical
\lcdm\ spectra (with $r = 0.1$) are shown for comparison. \\}
\label{fig:eebb}
\end{figure}

As suggested by \citet{bond00}, we apply a transformation
\begin{equation}
Z_b = \ln({\mathscr C}_b + x_b)
\end{equation}
to account for the fact that the probability of the true model value,
given an observed band power, is offset-lognormally distributed.  The
offsets $x_b$ describe the noise power spectra on the sky
(i.e., corrected for filter and beam bias) and are calculated from
simulations.  We calculate $x_b$ for the $TT$, $EE$, and $BB$ spectra,
but we assume Gaussian distributions for the $TE$, $TB$, and $EB$ band
powers since the values can be negative.  The \bicep\ band powers,
$x_b$ offsets, covariance matrices, and band power window functions
are available online at {\bf http://bicep.caltech.edu}.

\begin{figure*}
\begin{center}
\resizebox{\textwidth}{!}{\includegraphics{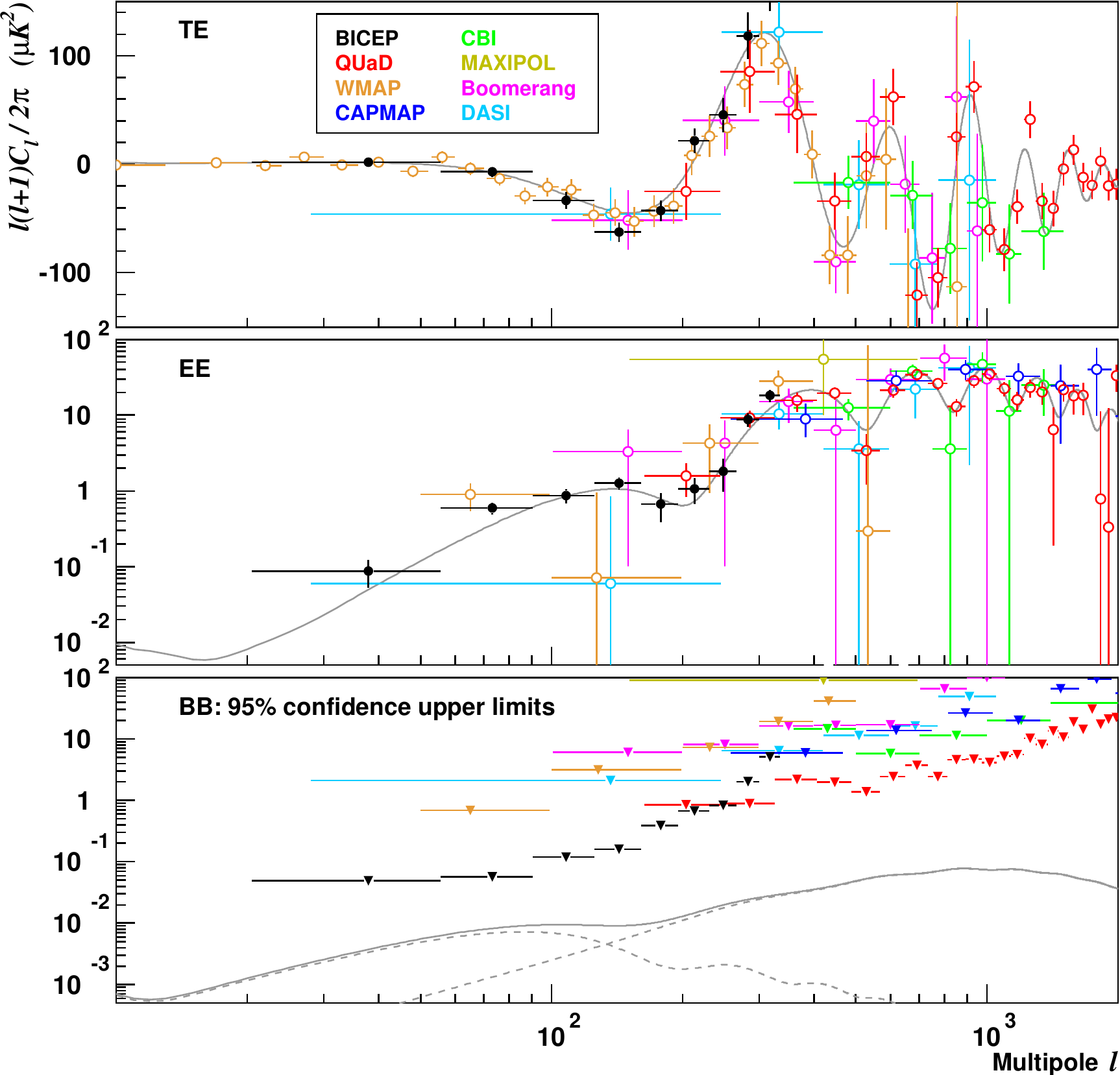}}
\caption{\bicep's $TE$, $EE$, and $BB$ power spectra complement
existing data from other CMB polarization
experiments~\citep{leitch05,montroy06,piacentini06,sievers07,wu07,
bischoff08,nolta09,brown09}.  Theoretical spectra from a \lcdm\ model 
with $r = 0.1$ are shown for comparison; the $BB$ curve is the sum of 
the inflationary and gravitational lensing components.  At degree angular
scales, \bicep's constraints on $BB$ are the most powerful to
date. \\}
\label{fig:pscomp}
\end{center}
\end{figure*}

Figure~\ref{fig:data_vs_lcdm} shows a comparison of the
frequency-combined spectra with a \lcdm\ model derived from \wmap\
five-year data.  The power spectrum results are confirmed by the
alternate analysis pipeline (open circles,
Figure~\ref{fig:data_vs_lcdm}).  \bicep\ contributes the first high 
S/N polarization measurements around $\ell\sim100$, as
illustrated by Figure~\ref{fig:eebb}, which shows the $EE$ peak at
$\ell\sim140$ in greater detail; the $BB$ spectrum is overplotted for
comparison.  Figure~\ref{fig:pscomp} shows \bicep's $TE$ and $EE$
spectra, as well as the 95\% confidence upper limits on $BB$, in
addition to other recent CMB polarization data.  To obtain the $BB$
upper limits, we apply offset-lognormal transformations to the band
powers and integrate the positive portion of the band power
probability distributions up to the 95\% point.  \bicep\ measures $EE$
in a multipole window that complements existing data from other
experiments, and all nine band powers have $>2\sigma$ significance.
The constraints on $BB$ are the most powerful to date.

\section{Consistency with \lcdm}\label{s:datalcdm}

The power spectra of the CMB are well described by a \lcdm\ model,
which, in its simplest form, has six parameters that have been
constrained by numerous experiments.  We check the consistency of the
\bicep\ band powers with this model by performing a $\chi^2$ test.  We
start by using \camb\ to calculate theoretical power spectra, using
\lcdm\ parameters derived from \wmap\ five-year data (and $r = 0$),
and we then compute expected band power values, ${\mathscr C}^X_b$, using
the band power window functions described in \S\ref{s:bpwf}.  Absolute 
gain and beam systematic errors ($G^X$ and $S_b$, as described in
\S\ref{s:gain_beam_sys}) are included by adding their contributions to
the band power covariance matrix, $M^X_{ab}$:
\begin{equation}
{\cal M}^X_{ab} = M^X_{ab} + (G^X)^2
{\mathscr C}^X_a {\mathscr C}^X_b + S_a S_b {\mathscr C}^X_a {\mathscr C}^X_b.
\end{equation}
The $S_b$ factors are formed from linear combinations of the four
frequencies (100~GHz auto, 150~GHz auto, 100$\times$150,
150$\times$100), using the weights described in \S\ref{s:combineps}.
Because $M^X_{ab}$ is obtained from a limited number of simulations,
the far off-diagonal terms are dominated by noise; we therefore use
only the main and first two off-diagonal terms of $M^X_{ab}$ in this
calculation. (We have tested that results are essentially unchanged
including one, two, or all off-diagonal terms.)  
For each power spectrum, the observed and theoretical band
powers are compared by evaluating
\begin{equation}
\chi^2 = [\hat{\pmb{\mathscr C}}^X-{\pmb{\mathscr C}}^X]^\top ({\pmb{\cal M}}^X)^{-1} [\hat{\pmb{\mathscr C}}^X-{\pmb{\mathscr C}}^X]
\label{eq:chi2_datalcdm}
\end{equation}
over the nine bins that span $21 \leq \ell \leq 335$.  In the case of
the $TT$, $EE$, and $BB$ spectra, offset-lognormal transformations,
\begin{eqnarray}
{\hat Z}^X_b &=& \ln ({\hat{\mathscr C}}^X_b + x^X_b), \label{eq:ol1} \\
Z^X_b &=& \ln ({\mathscr C}^X_b + x^X_b), \label{eq:ol2} \\
(D^X_{ab})^{-1} &=& ({\cal M}^X_{ab})^{-1} ({\hat{\mathscr C}}^X_a +
x^X_a) ({\hat{\mathscr C}}^X_b + x^X_b), \label{eq:ol3}
\end{eqnarray}
are applied to the data, expected band powers, and inverse covariance
matrix, and $\chi^2$ is calculated using the transformed quantities.

We perform the same calculations for a set of 500 signal-plus-noise
simulations, and the simulated $\chi^2$ distributions are used to
determine the probabilities to exceed the $\chi^2$ values of the data.
The $\chi^2$ and PTE values are listed in
Figure~\ref{fig:data_vs_lcdm}, which shows a comparison of our data
with the \lcdm\ model.  The \bicep\ data are consistent with \lcdm,
and this result is confirmed by the alternate analysis pipeline.

\section{Constraint on Tensor-to-Scalar Ratio from $BB$}
\label{s:t2s}

\begin{figure*}
\begin{center}
\resizebox{\textwidth}{!}{\includegraphics{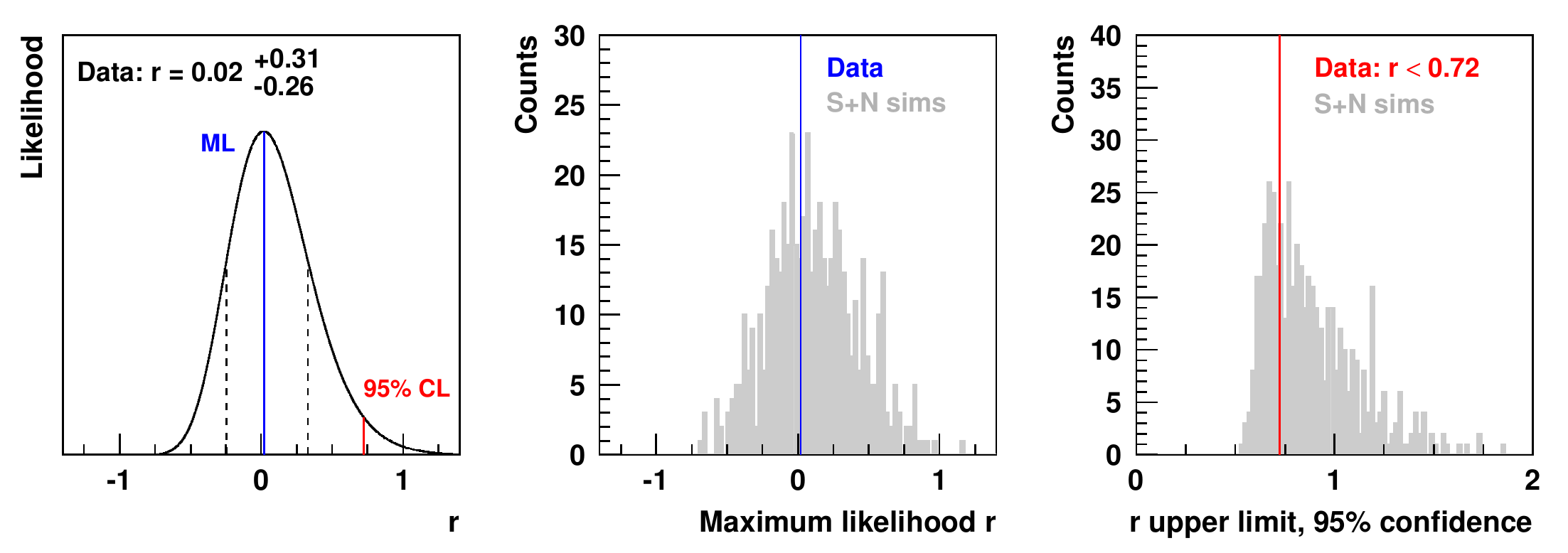}}
\caption{$r$ likelihood function calculated from the \bicep\ $BB$
spectrum is shown in the left panel and peaks at a value of $r =
0.02$.  For comparison, a histogram of maximum likelihood $r$ values
derived from 500 signal-plus-noise simulations (with $r=0$ input) is
shown in the central panel.  In the right panel, we derive 95\%
confidence upper limits on $r$ from the both the simulated and real
likelihoods.  The data yield an upper limit of $r < 0.72$, which lies
within the simulated distribution. \\}
\label{fig:r}
\end{center}
\end{figure*}

\bicep\ was designed with the goal of measuring the $BB$ spectrum at
degree angular scales in order to constrain the tensor-to-scalar ratio
$r$.  We define $r = \Delta^2_h(k_0) / \Delta^2_R(k_0)$, where
$\Delta^2_h$ is the amplitude of primordial gravitational waves,
$\Delta^2_R$ is the amplitude of curvature perturbations, and we
choose a pivot point $k_0 = 0.002$~Mpc$^{-1}$.  The 
tightest published upper 
limit is $r<0.22$ at 95\% confidence and is derived
from a combination of the \wmap\ five-year measurements of the $TT$
power spectrum at low $\ell$ 
with measurements of Type~Ia supernovae and baryon acoustic
oscillations~\citep{komatsu09}.  

As a
method for constraining $r$, a direct measurement of the $BB$ spectrum
has two advantages.  First, measurements of $TT$ are ultimately
limited by cosmic variance at large angular scales, and the
temperature data from \wmap\ have already reached that limit.  Second,
$r$ constraints from $TT$ are limited by parameter degeneracies; in
particular, there is a strong degeneracy with the scalar spectral
index $n_s$.
The inflationary $BB$ spectrum, in contrast, suffers little from
parameter degeneracies at the angular scales that \bicep\ probes, and
the $BB$ amplitude depends primarily on $r$.

To determine a constraint on $r$ from \bicep's $BB$ spectrum, we
compare the measured band powers, ${\hat{\mathscr C}}^{BB}_b$, to a
template $BB$ curve and vary the amplitude of the inflationary component, assuming
that it simply scales with $r$.  The template $BB$ curves are calculated using fixed
\lcdm\ parameters derived from \wmap\ five-year data and a wide range
of trial $r$ values.  We include a constant $BB$ component from gravitational 
lensing, although the contribution is negligible at low multipoles.  
From the template $BB$ curves, we compute the expected
band powers, ${\mathscr C}^{BB}_b(r)$.  We apply offset-lognormal
transformations to the data, expected band powers, and covariance
matrix (Equations~\ref{eq:ol1}--\ref{eq:ol3}), and we calculate
\begin{equation}
\chi^2(r) = [{\hat {\bf Z}}^{BB}-{\bf Z}^{BB}(r)]^\top [{\bf D}^{BB}(r)]^{-1} [{\hat {\bf Z}}^{BB}-{\bf Z}^{BB}(r)]
\label{eq:r_chi2}
\end{equation}
at each trial $r$ value.  The likelihood is then
\begin{equation}
{\cal L} \propto \frac{1}{\sqrt{\det[{\bf D}^{BB}(r)]}} ~e^{-\chi^2(r)/2}.
\label{eq:r_like}
\end{equation}
Figure~\ref{fig:r} (left panel) shows that the maximum likelihood
value obtained from \bicep\ data is $r = 0.02^{+0.31}_{-0.26}$.  For
comparison, the central panel shows the maximum likelihood $r$ values
from 500 signal-plus-noise simulations with $r = 0$ input spectra.  We
calculate the upper limit on $r$ by integrating the positive portion
of the likelihood up to the 95\% point, and we find that the \bicep\
$BB$ spectrum constrains $r<0.72$ at 95\% confidence.  This constraint
is consistent with those obtained from simulations
(Figure~\ref{fig:r}, right panel).

\begin{figure}[t]
\begin{center}
\resizebox{0.7\columnwidth}{!}{\includegraphics{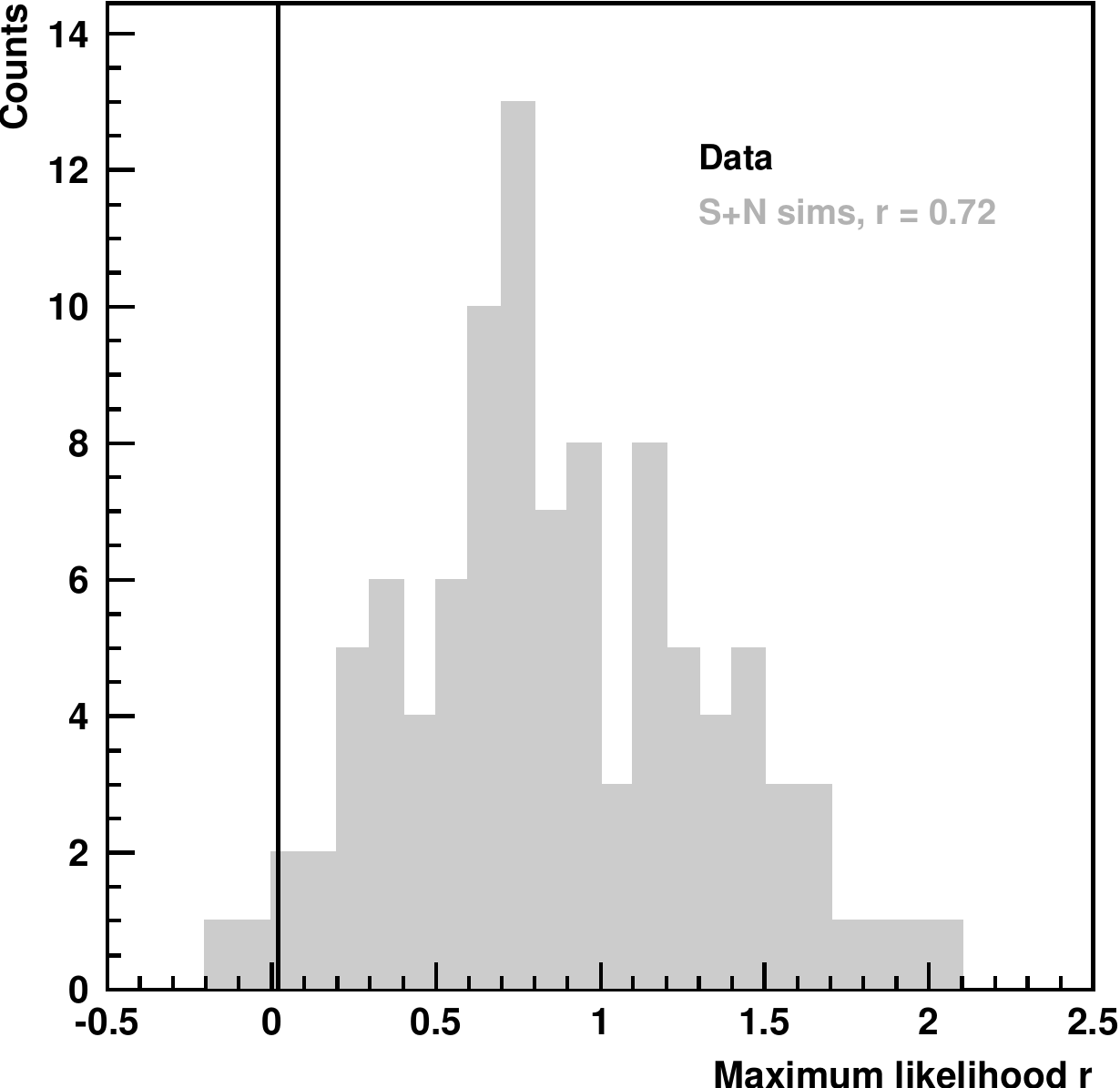}}
\caption{Maximum likelihood $r$ values are shown for 100
signal-plus-noise simulations with input spectra for $r = 0.72$, a value
chosen to match \bicep's 95\% upper limit.
As expected, the distribution peaks around 0.72, showing the \bicep\ pipeline
recovers an unbiased estimate of $r$.  Only two of the 100 simulations fall below
$r = 0.02$, the maximum likelihood value derived from \bicep\ data,
offering an alternative demonstration that $r = 0.72$ is strongly disfavored
by the \bicep\ data. \\}
\label{fig:r0p7}
\end{center}
\end{figure}

We have cross-checked the \bicep\ $r$ constraint with two methods.
First, the alternate analysis pipeline yields very similar estimates
of $r$, adding confidence to our result.  Second, we have generated
100 signal-plus-noise simulations using input spectra corresponding to
$r = 0.72$, and we have calculated $r$ likelihood curves for each of
these realizations.  Figure~\ref{fig:r0p7} shows the histogram of
maximum likelihood $r$ values, which peaks as expected around the input value of
0.72, confirming our pipeline recovers an unbiased estimate of $r$.
Only two of 100 simulations have maximum likelihood $r$ values that
fall below the data; this simple alternative statistic to the Bayesian
95\% upper limit suggests at a similar level of confidence that the \bicep\ data
excludes the $r=0.72$ hypothesis.

While limits on $r$ derived from CMB data are still driven by the
\wmap\ measurement of the $TT$ power spectrum, \bicep's limits on the
amplitude of the $BB$ spectrum are an order of magnitude more powerful
than any previous measurement.  The improvement in the power of $BB$
to constrain $r$ is illustrated by repeating the above analysis using
\wmap\ $BB$ data, where we obtain a limit of $r<6$, versus the \bicep\
constraint of $r<0.72$.

\section{Conclusions}

Motivated by the exciting possibility of detecting, albeit indirectly,
the gravitational wave background due to inflation, many efforts are
underway to develop the instrumentation and methods necessary to
search for the \bmode\ component of CMB polarization at degree angular
scales.  In this paper, we have presented initial results from \bicep,
the first experiment optimized specifically to search for the
inflationary \bmode\ signal.  Analysis of a subset of the first two
years of observations provides the most sensitive measurement to date
of CMB polarization at degree angular scales.  The $TT$, $TE$, and
$EE$ angular power spectra are measured with high S/N; the
$TB$, $BB$, and $EB$ spectra are consistent with zero.  The spectra
are consistent with a \lcdm\ model using parameters derived from
\wmap\ five-year data.  The polarization data pass all of the
statistical tests that we have been able to devise.  Furthermore, the
results have been cross-checked by two independent analysis pipelines,
whose agreement is excellent at all stages of the data processing,
including maps, power spectra, jackknives, and comparisons with
theoretical models.  A detailed study of \bicep's potential systematic
errors indicate that all systematic effects are below the level of the
statistical errors and are not limiting factors in the analysis.

\bicep\ has detected, for the first time, the first peak in the
angular power spectrum of the \emode\ polarization.  The combination
of these results from \bicep\ and of results from its sister
experiment, \QUAD, confirm with significant precision the theoretical
prediction for the shape and amplitude of the \emode\ spectrum on
angular scales that span its first six peaks.  In addition, both
experiments have shown that there is no \bmode\ component comparable
in amplitude to the \emode\ component.

Because \bicep\ is designed specifically to probe the degree angular
scales at which the inflationary \bmode\ signature peaks, \bicep's
upper limits on the \bmode\ angular power spectrum provide the first
meaningful constraint on the inflationary gravitational wave
background to come directly from CMB \bmode\ polarization, $r =
0.02^{+0.31}_{-0.26}$ or $r < 0.72$ at 95\% confidence.  Because this
limit is dominated by statistical noise,    
constraints on $r$ may be expected to improve in proportion to 
observing time (not its square root), and 
it is likely that analysis of
the remaining data from the three seasons of observing will provide
improved limits on $r$.  Future \bicep\ analyses will include the full
data set, explore relaxing the conservative cuts employed for the
initial analysis, and optimize the power spectrum recovery, all of
which will strengthen \bicep\ constraints on $r$ significantly.

\acknowledgements

We dedicate this work to the memory of Andrew Lange, whose tragic and
untimely death has deeply pained us all.  He was and always will be an
inspiration to us, and he is immeasurably missed.

\bicep\ is supported by NSF Grant No. OPP-0230438, Caltech President's
Discovery Fund, Caltech President's Fund PF-471, JPL Research and
Technology Development Fund, and the late J.~Robinson.  We thank the
South Pole Station staff for helping make our observing seasons a
success.  We also thank Joanna Dunkley, Nathan Miller, and our
colleagues in \acbar, \boom, \QUAD, \bolocam, \spt, and \wmap\ for
advice and helpful discussions, and Kathy Deniston for logistical and
administrative support.  We gratefully acknowledge support of
individual team members by the NASA Graduate Fellowship program (H.C.C.),
NSF PECASE Award No.\ AST-0548262 (B.G.K.), the John~B. and Nelly Kilroy
Foundation (J.M.K.), the U.S. DOE contract to SLAC No.\ DE-AC02-76SF00515
(C.L.K. and J.E.T.), KICP (C.P. and C.S.), and the NASA Science Mission
Directorate via the US Planck Project (G.R.).

\bibliographystyle{apj}
\bibliography{2008_bicep_cmb_2yr}

\end{document}

%% file: defs.tex
\def\bicep{{\sc Bicep}}
\def\bolocam{{\sc Bolocam}}
\def\dasi{{\sc Dasi}}

\def\acbar{{\sc Acbar}}
\def\QUAD{{\sc QUaD}}

\def\boom{{\sc Boomerang}}
\def\wmap{{\sc Wmap}}
\def\spt{{\sc SPT}}

\def\camb{{\tt CAMB}}

\def\anafast{{\tt anafast}}
\def\synfast{{\tt synfast}}
\def\healpix{{\tt Healpix}}

\def\spice{{\tt Spice}}

\def\muK{~\mu{\rm K}}

\def\deg{^\circ}
\def\emode{$E$-mode}
\def\bmode{$B$-mode}

\def\lcdm{$\Lambda$CDM}
\def\grad{{\vec \nabla}}



%% file: ptetab.tex
Scan direction \\
$EE$ & 0.532 & 0.588 & 0.740 \\
$BB$ & 0.640 & 0.568 & 0.212 \\
$EB$ & 0.816 & 0.962 & 0.924 & 0.358 \\
\\
Elevation coverage \\
$EE$ & 0.576 & 0.546 & 0.924 \\
$BB$ & 0.584 & 0.288 & 0.618 \\
$EB$ & 0.872 & 0.728 & 0.892 & 0.892 \\
\\
Boresight angle \\
$EE$ & 0.916 & 0.448 & 0.320 \\
$BB$ & 0.242 & 0.548 & 0.592 \\
$EB$ & 0.912 & 0.100 & 0.392 & 0.944 \\
\\
Temporal split \\
$EE$ & 0.378 & 0.208 & 0.796 \\
$BB$ & 0.788 & 0.020 & 0.852 \\
$EB$ & 0.370 & 0.580 & 0.476 & 0.232 \\
\\
Season split \\
$EE$ & 0.564 & 0.716 & 0.216 \\
$BB$ & 0.790 & 0.992 & 0.056 \\
$EB$ & 0.806 & 0.514 & 0.456 & 0.986 \\
\\
Focal plane $QU$ \\
$EE$ & 0.670 & 0.014 & 0.994 \\
$BB$ & 0.896 & 0.804 & 0.576 \\
$EB$ & 0.236 & 0.806 & 0.234 & 0.560 \\